%% file: pilot_L0.tex
\documentclass[a4paper,traditabstract,longauth]{aa} 
\usepackage{amsmath}
\usepackage{amsfonts}
\usepackage{amssymb}

\usepackage{graphicx}
\usepackage{color}
\usepackage{natbib}
\usepackage{longtable,lscape}
\usepackage{txfonts}
\usepackage{newlfont}
\usepackage{textcomp}
\usepackage{times}
\usepackage{units}
\usepackage{subcaption}

\newcommand{\bc}{\begin{center}}
\newcommand{\ec}{\end{center}}
\newcommand{\bi}{\begin{itemize}}
\newcommand{\ei}{\end{itemize}}
\newcommand{\ben}{\begin{enumerate}}
\newcommand{\een}{\end{enumerate}}

\newcommand{\Planck}{\it Planck}
\newcommand{\Herschel}{\it Herschel}
\newcommand{\PILOT}{\it PILOT}
\newcommand{\Pilot}{\it PILOT}
\newcommand{\Artemis}{\it ArT\'eMiS}
\newcommand{\APEX}{\it APEX}
\newcommand{\BICEP}{\it BICEP}

\newcommand{\CNES}{\it CNES}

\newcommand{\CMZ}{CMZ}

\newcommand{\micron}{$\mu$m}
\newcommand{\mic}{\,\mu\mathrm{m}}

\def\ts {\thinspace}
\def\kms {\ifmmode{\ts \mathrm{km}/\mathrm{s}}\else{\ts km/s}\fi}
\def\cc {\ifmmode{{\rm \ts cm}^{-2}}\else{\ts cm$^{-2}$}\fi}
\def\mo {\ifmmode{{\rm M}_{\odot}}\else{M$_{\odot}$}\fi}
\def\msol {\ifmmode{{\rm M}_{\odot}}\else{M$_{\odot}$}\fi}
\def\lsol {\ifmmode{{\rm L}_{\odot}}\else{L$_{\odot}$}\fi}
\def\mpcsq {\ifmmode{{\rm M}_{\odot}\ts {\rm pc}^{-2}}\else{M$_{\odot}$}\ts pc$^{-2}$\fi}

\newcommand{\scanamorphos}{\sc Scanamorphos}
\newcommand{\Scanamorphos}{\sc Scanamorphos}
\newcommand{\ROMA}{\sc Roma}
\newcommand{\IAU}{IAU}
\newcommand{\COSMO}{COSMO}

\newcommand{\Lzero}{L0}

\newcommand{\Estadius}{\it Estadius}
\newcommand{\scene}{observing tile}
\newcommand{\scenes}{observing tiles}
\newcommand{\HWP}{HWP}
\newcommand{\ICS}{ICS}

\newcommand{\FWHM}{FWHM}

\newcommand{\TRANS}{TRANS}
\newcommand{\REFLEX}{REFLEX}
\newcommand{\NIKAdeux}{NIKA2}
\newcommand{\GC}{Galactic center}
\newcommand{\Sfunction}{angle dispersion function}  

\newcommand{\lag}{\delta} 

\newcommand{\Bfield}{\vec{B}}

\newcommand{\Bperp}{\langle\Bfield_\perp\rangle}         

\newcommand{\Resp}{R} 
\newcommand{\Ratmo}{R^\textrm{atmo}}  
\newcommand{\glon}{\ell}          
\newcommand{\glat}{\textrm{b}}          

\usepackage{mathrsfs}
\usepackage{mathbbol}

\definecolor{purple1}{rgb}{0.4,0.,0.6}

\definecolor{orange}{rgb}{1.0, 0.4, 0.0}

\include{journals_defs}
\include{polar_definitions}

\begin{document}


\title{The geometry of the magnetic field in the Central Molecular Zone measured by PILOT}

\author{
A.~Mangilli\inst{\ref{IRAP}}\and
J.~Aumont\inst{\ref{IRAP}}\and
J.-Ph.~Bernard\inst{\ref{IRAP}}\and
A.~Buzzelli\inst{\ref{ROMA1}}\and
G.~de~Gasperis\inst{\ref{ROMA1}}\and
J.B.~Durrive\inst{\ref{IRAP}}\and
K.~Ferriere\inst{\ref{IRAP}}\and
G.~Fo\"enard\inst{\ref{IRAP}}\and
A.~Hughes\inst{\ref{IRAP}}\and
A.~Lacourt\inst{\ref{IRAP}}\and
R.~Misawa\inst{\ref{IRAP}}\and
L.~Montier\inst{\ref{IRAP}}\and
B.~Mot\inst{\ref{IRAP}}\and
I.~Ristorcelli\inst{\ref{IRAP}}\and
H.~Roussel\inst{\ref{IAP}}\and
P.~Ade\inst{\ref{Cardiff}}\and
D.~Alina\inst{\ref{Astana}}\and
P.~de~Bernardis\inst{\ref{ROMA2}}\and
E.~de Gouveia Dal Pino\inst{\ref{SaoPaolo}}\and
J.P.~Dubois\inst{\ref{IAS}}\and
C. Engel\inst{\ref{IRAP}}\and
P.~Hargrave\inst{\ref{Cardiff}}\and
R.~Laureijs\inst{\ref{ESTEC}}\and
Y.~Longval\inst{\ref{IAS}}\and
B.~Maffei\inst{\ref{IAS}}\and
A.M.~Magalhães\inst{\ref{SaoPaolo}}\and
C. Marty\inst{\ref{IRAP}}\and
S.~Masi\inst{\ref{ROMA2}}\and
J.~Montel\inst{\ref{CNES}}\and
F.~Pajot\inst{\ref{IRAP}}\and
L.~Rodriguez\inst{\ref{CEA}}\and
M.~Salatino\inst{\ref{Stanford}}\and
M.~Saccoccio\inst{\ref{CNES}}\and
S.~Stever\inst{\ref{IAS}}\and
J.~Tauber\inst{\ref{ESTEC}}\and
C.~Tibbs\inst{\ref{ESTEC}}\and
C.~Tucker\inst{\ref{Cardiff}}
}

\institute{
IRAP, Universit\'e de Toulouse, CNRS, CNES, UPS, (Toulouse), France\label{IRAP}\and
Dipartimento di Fisica, Universit\`a di Roma ``Tor Vergata'', via della Ricerca Scientifica 1, 00133 Roma, Italy\label{ROMA1}\and
Institut d’Astrophysique de Paris, Universit\'e Pierre et Marie Curie (UPMC), Sorbonne Universit\'e, CNRS (UMR 7095), 75014 Paris, France\label{IAP}\and
Department of Physics and Astrophysics, PO BOX 913, Cardiff University, 5 the Parade, Cardiff, UK\label{Cardiff}\and
Department of Physics, School of Science and Technology, Nazarbayev University, Astana 010000, Kazakhstan\label{Astana}\and
Universit\`a degli studi di Roma ``La Sapienza'', Dipartimento di Fisica, P.le A. Moro, 2, 00185, Roma, Italia\label{ROMA2}\and
Centre National des Etudes Spatiales, DCT/BL/NB, 18 Av. E. Belin, 31401 Toulouse, France\label{CNES}\and
Instituto de Astronomia, Geof\'isica e Ci\^encias Atmosf\'ericas - Universidade de S\~ao Paulo (IAG-USP), Rua do Mat\~ao, S\~ao Paulo, 05508-090, Brasil\label{SaoPaolo}\and
Institut d’Astrophysique Spatiale (IAS), B\^at 131, Universit\'e Paris XI, Orsay, France\label{IAS}\and
Scientific Support Office, SRE-S, ESTEC, PO Box 299,2200AG Noordwijk, The Netherlands\label{ESTEC}\and
Thales Services, Toulouse, France\label{Thales}\and
CEA/Saclay, 91191 Gif-sur-Yvette Cedex, France\label{CEA}\and
Department of Physics, Stanford University, Stanford, California 94305, USA\label{Stanford}
}

\abstract{We present the first far infrared (FIR) dust emission polarization map covering the full extent Milky Way's Central molecular zone ({\CMZ}). The data, obtained with the {\PILOT} balloon-borne experiment, covers the Galactic Center region $-2\,\degr<\glon<2\,\degr$, $-4\,\degr<\glat<3\,\degr$ at a wavelength of 240\,\micron\ and an angular resolution $2.2\,\arcmin$. 
From our measured dust polarization angles, we infer a magnetic field orientation projected onto the plane of the sky that is remarkably ordered over the full extent of the CMZ, with an average tilt angle of $\simeq 22\,\degr$ clockwise with respect to the Galactic plane. Our results confirm previous claims  that the field traced by dust polarized emission is oriented nearly orthogonal to the field traced by GHz radio synchrotron emission in the Galactic Center region. 
The observed field structure is globally compatible with the latest {\Planck} polarization data at 353\,GHz and 217\,GHz. Upon subtraction of the extended emission in our data, the mean field orientation that we obtain shows good agreement with the mean field orientation measured at higher angular resolution by the JCMT within the 20\,\kms\ and 50\,\kms\ molecular clouds. 
We find no evidence that the magnetic field orientation is related to the 100\,pc twisted ring structure within the CMZ. We propose that the low polarization fraction in the Galactic Center region and the highly ordered projected field orientation can be reconciled if the field is strong, with a 3D geometry that is is mostly oriented $\simeq 15\,\degr$ with respect to the line-of-sight towards the Galactic center.  
Assuming equipartition between the magnetic pressure and ram pressure, we obtain magnetic field strengths estimates as high as a few mG for several CMZ molecular clouds.
}

\authorrunning{PILOT Collaboration}
\titlerunning{Galactic center polarization with PILOT}

\maketitle
\section{Introduction}
\label{sec:Intro}

The interstellar medium (ISM) near the Galactic center (GC)
is dominated by the
Central Molecular Zone (CMZ), a large reservoir of dense molecular gas with a total mass of $\sim10^{7}$\,\msol\ and typical gas densities of a few times $10^{4}$\,\cc\ \citep[e.g,][]{Ferriere2007}.
The CMZ is structured into a thin, elliptical sheet of gas that is oriented roughly parallel to the Galactic plane. In the plane of the sky, it extends out to $r \sim 250$\,pc at positive longitudes and $r \sim 150$\,pc at negative longitudes, with a FWHM thickness $\sim 30$\,pc \citep[e.g.,][]{heiligman_87, bally&swh_88}.\footnote{ We assume that the Sun lies at a distance of 8.5\,kpc from Sgr~A$^*$, the bright and compact radio source at the dynamical center of the Galaxy. Accordingly, an angular distance of $1\,^\circ$ in the plane of the sky corresponds to a linear distance of approximately 150\,pc near the GC.}
The CMZ itself contains a ring-like feature with mean radius $\sim 180$\,pc, and, deeper inside, a population of dense molecular clouds \citep[e.g.,][]{bally&swh_88, sofue_95a, sofue_95b}. These clouds appear to be arranged along a twisted elliptical ring \citep{Molinari_etal2011}. 

\noindent The first observational clues to the orientation of the interstellar magnetic field in the CMZ
date back to the 1980s, when radio astronomers discovered systems of radio continuum filaments running nearly perpendicular to the Galactic plane \citep{yusef&mc_84, liszt_85}.
As summarized by \cite{morris_96}, these filaments are typically a few to a few tens of parsecs long and a fraction of a parsec wide. They appear straight or mildly curved along their entire length. Their radio continuum emission is linearly polarized and has a spectral index consistent with synchrotron radiation, leading to their denomination as non-thermal radio filaments (NRFs or NTFs).

The long and thin shape of NRFs strongly suggests that they follow magnetic field lines. This suggested alignment is confirmed by the measured radio polarization angles (corrected for Faraday rotation), which indicate that, in the plane of the sky, the magnetic field inside NRFs is indeed oriented along their long axes \citep[e.g.,][]{tsuboi&ihtk_85, tsuboi&iht_86, reich_94, lang&akl_99}. From this, it has been concluded that the magnetic field in the CMZ is approximately vertical (i.e., perpendicular to the Galactic plane), at least close to the plane. At larger distances from the plane, the orientation of NRFs tends to lean outwards (i.e., away from the vertical), consistent with the magnetic field having an overall poloidal geometry \citep{morris_90}.

\noindent Following the initial discovery of the NRF phenomenon, many new NRFs have been identified in the CMZ. By plotting the sky distribution of all the (confirmed and likely candidate) NRFs detected at 20\,cm, \cite{yusef&hc_04} observed that only the longer NRFs are nearly straight and aligned with the vertical; the shorter NRFs exhibit a broad range of orientations, with only a loose trend toward the vertical. This could indicate that the magnetic field in the CMZ is not as ordered as initially claimed, and that it has a significant fluctuating component.

\noindent Faraday rotation measurements (RMs) toward NRFs provide valuable information on the magnetic field in the diffuse ionized medium near the GC. \cite{Novak_etal2003} collected all the available RMs toward NRFs within $1\,^\circ$ ($\simeq 150$\,pc) of Sgr~A$^*$ and, by doing so, they showed that there is a clear pattern in the sign of RMs, such that RM~$ > 0$ in the North-East and South-West quadrants and RM~$ < 0$ in the North-West and South-East quadrants. These authors argued that the observed RM pattern could be explained by the model of \cite{Uchida_etal1985}, in which an initially vertical magnetic field
is sheared out in the azimuthal direction by the differential rotation of the dense clouds present near the Galactic plane.

\noindent A different RM pattern was obtained by \cite{roy&rs_05}, who measured the RMs of 60 background extragalactic sources through a $12\,^\circ \times 4\,^\circ$ window centered on Sgr~A$^*$ and found mostly positive values, with no evidence for a sign reversal either across the rotation axis or across the midplane. \cite{roy&rs_08} pointed out that this RM distribution is consistent with either the large-scale Galactic magnetic field having a bisymmetric spiral configuration or the magnetic field in the central region of the Galaxy being oriented along the Galactic bar.

\noindent More relevant to the present paper, far-infrared (FIR) and submillimeter (submm) polarization studies of dust thermal emission make it possible to probe the orientation, in the plane of the sky, of the magnetic field inside dense molecular clouds.

\noindent FIR polarimetry of the innermost ($\sim 3$\,pc diameter) GC region was first performed by \cite{Werner_etal1988}, who detected linear polarization of $100\,\mic$ emission from three locations in the inner circumnuclear ring (CNR).
At each location, they measured polarization angles $\simeq 90\,^\circ - 100\,^\circ$, implying a mean magnetic field oriented at $\simeq 0\,^\circ - 10\,^\circ$ East of North
, i.e., within $\simeq 10\,^\circ - 20\,^\circ$ of the plane of the CNR.
Since the CNR is thought to be in differential rotation,
they concluded that its mean magnetic field must be predominantly azimuthal (with respect to its rotation axis). Finally, from the measured polarization fractions, they inferred that the magnetic field has a turbulent component comparable to the mean azimuthal field.

\noindent As a follow-up to the work of \cite{Werner_etal1988}, \cite{hildebrand&gpw_90} measured the linear polarization of the $100\,\mic$ emission at six positions in the CNR, along the main FIR emission ridge, and they confirmed that the mean magnetic field is approximately parallel to the plane of the CNR. \cite{hildebrand&ddf_93} then expanded the set of $100~\mu$m polarization measurements, to cover not only the main FIR emission ridge, but the whole area of the CNR. A few of their observed positions happen to fall inside the area of the Northern streamer. By analyzing the corresponding polarization measurements, \cite{hildebrand&d_94} found that the magnetic field runs parallel to the axis of the streamer along its entire length, i.e., out to a projected distance $\simeq 4$\,pc from Sgr~A$^*$.

\noindent The first submm polarimetric observations of the GC region were carried out by \cite{Novak_etal2000}. They targeted three separate 5\,pc~$\times$~5\,pc areas, centered on the CNR and on the peaks of the M$-$0.02$-$0.07 and M$-$0.13$-$0.08 molecular clouds, respectively, and in each area they detected linear polarization of the $350\,\mic$ emission.
For the CNR, they inferred a mean magnetic field orientation that is approximately north-south, in good overall agreement with the FIR results. In M$-$0.02$-$0.07, the field appears to have two distinct behaviors: upstream of the Sgr~A East shock, it runs nearly perpendicular to the Galactic plane, while downstream it closely follows the curved ridge of shock-compressed material. In M$-$0.13$-$0.08, the field is on average parallel to the long axis of the cloud,
with a spiky structure toward the CNR, which suggests that the field has been stretched out by the tidal forces that gave the cloud its elongated shape.

\noindent \cite{Novak_etal2003} observed a much larger, 170\,pc~$\times$~30\,pc, area around Sgr~A$^*$.
Their $450\,\mic$ polarization map clearly shows that the magnetic field threading molecular clouds is, on the whole, approximately parallel to the Galactic plane. To reconcile the horizontal field measured in molecular clouds with the poloidal field traced by the NRFs, \cite{Novak_etal2003} proposed that the large-scale magnetic field in the GC region is predominantly poloidal in the diffuse ISM and predominantly toroidal in dense regions along the Galactic plane, where it was sheared out in the azimuthal direction by the differential rotation of the dense gas.

\noindent \cite{Chuss_etal2003} performed additional $350\,\mic$ polarimetric observations toward 8 selected areas within the central 50\,pc. These  measurements, combined with those of \cite{Novak_etal2000}, offer a broader and more complete view of the magnetic field morphology in the GC region, which conveys the general impression that the field is globally organized on scales much larger than the typical sizes of molecular clouds, while also being subject to strong local distortions by environmental forces such as shock compression and tidal shearing. \cite{Chuss_etal2003} also found that the measured magnetic field orientation depends on the molecular gas density, being generally parallel to the Galactic plane in high-density regions and generally perpendicular to it in low-density regions.

\noindent Near-infrared (NIR) polarization observations of starlight extinction by dust also offer a promising tool to trace the magnetic field orientation in dense regions near the GC. \cite{nishiyama&thk_09} obtained a NIR polarization map of a 50\,pc~$\times$~50\,pc area centered on Sgr~A$^*$. Compared to earlier NIR polarimetric observations toward the GC \citep[e.g.,][]{eckart&ghs_95, ott&eg_99}, they were able, for the first time, to separate out the contribution from foreground dust and to isolate the polarization arising within $\sim (1-2)$\,kpc of Sgr~A$^*$. They inferred that the distribution of polarization angles exhibits a strong peak in a direction nearly parallel to the Galactic plane, in good agreement with the results of FIR/submm polarimetry. However, in contrast to \cite{Chuss_etal2003}, \cite{nishiyama&thk_09} found no indication that the magnetic field orientation depends on gas density -- the field appears to be everywhere horizontal, including in the diffuse ISM.

\noindent All sky maps of the thermal dust and synchrotron polarized emission have been obtained with the {\Planck} satellite at wavelengths above 850\,\micron\ (353\,GHz) and angular resolution above $5\,\arcmin$. These maps obviously include the {\GC} region, but no specific study of this region was discussed in the literature. The analysis of the all-sky map of thermal dust polarization at 850\,\micron\ \citep{Planck2015_intermediate_XIX} has clearly shown the existence of large variations of the polarization fraction from unexpectedly large values (up to $\polfrac\simeq20\,\%$) occurring essentially in the diffuse ISM, down to very low values occurring mostly on line-of-sights (LOS) with larger column densities. Over most of the sky included in the analysis, which avoided the Galactic plane (usually $|b|<5\,\degr$), the variations of $\polfrac$ appeared tightly anti-correlated with the polarization angle dispersion function ($\DeltaAng$) which measures the rotation of the observed field around a given sky location. This indicates that the intrinsically large dust polarization fraction is efficiently modulated by the $B$-field twisted geometry resulting in efficient depolarization of the signal both in the beam and along the LOS. \cite{Planck2015_intermediate_XX} showed that this behaviour is expected from MHD simulations of the ISM. These properties were confirmed in the latest analysis of the all-sky {\Planck} data at 353\,GHz based on the latest {\Planck} data release in \cite{Planck_polar_2018}, which also studied departures from the $\polfrac-\DeltaAng$ anti-correlation in details. They showed that the departures are unlikely to reflect large variations of the dust alignment efficiency up to a column density of $N_H=2\times10^{22}\,\mathrm{cm}^{-2}$. However, this analysis excluded regions near the Galactic plane ($|b|<2\,\degr$), and therefore the {\GC} discussed here. The main reason for avoiding the Galactic plane and in particular the inner Galactic regions was to avoid very long line-of-sight through the plane were many $B$-field reversal are likely to happen, which could have affected the trend with column density. An additional difficulty comes from specific systematic effects which could be affecting data near the plane, such as band-pass mismatch and/or contamination by molecular lines (such as the CO lines) in the {\Planck} photometric bands.

\noindent In this context, the {\Pilot} data bring useful information regarding both the structure of the magnetic field and the physics of dust. First, the angular resolution of the instrument ($2.2\arcmin$) is higher than that of the {\Planck} data, which helps following the field structure into small and/or distant objects, while the large instantaneous field-of-view of the instrument allows mapping large regions, allowing to bridge high resolution ground-based observation which often suffer from a limited sky coverage. Second, the data obtained with the {\Pilot} instrument operating at $240\mic$ will allow, in conjunction with polarization data obtained at longer wavelength, to constrain the polarization SED of thermal dust with a large lever arm, which is an important constraint for dust models. Finally, the systematic effects affecting the {\Pilot} data are likely to be of a different nature than that relevant for other instruments, which will help the analysis and interpretation of the polarized signals when co-analysis is possible.

\noindent In this paper, we present the map of polarized dust emission in the CMZ obtained at 240\,\micron\ by the {\PILOT} balloon-borne experiment. The {\PILOT} observations cover a $4 \times 6$\,degree field centred at $(l,b)=(0,0)$ (we refer to this observed field as ``\Lzero'' in the rest of the article), a much wider field than all previous  observations of polarized FIR emission in the GC region. The {\PILOT} maps reveal the projected magnetic field structure across the whole CMZ with an angular resolution of $2\,\arcmin2$, i.e. with $2.3$ times better linear resolution or $5$ times more spatial information than the 353\,GHz polarization data obtained by {\Planck}. 

\noindent The paper is organized as follows. In Sect.~\ref{sec:obs}, we describe the observing strategy. The {\PILOT} data processing is described in Sect.~\ref{sec:Data_processing}, including an overview of how we construct Stokes $\StokesI$, $\StokesQ$ and $\StokesU$ maps by using two map-making  software packages ({\ROMA} and {\Scanamorphos}) in Sect.~\ref{Sec:Map_making}. In Sect.~\ref{Sec:Polarization_results} we present the polarization results focusing on the polarization angles, we discuss the polarization maps in Sect.~\ref{sec:pilot_maps}, in  Sect.~\ref{Sec:mapmaking_comparison}, we validate them by comparing the results of the two independent map-making procedures and in Sect.~\ref{sec:comparison_planck} we compare our measurements with the ones obtained with the {\Planck} satellite. 
We discuss the results in Sect.~\ref{sec:discussion}, focusing on the relevance of the CMZ in Sect.~\ref{sec:cmz}, on the magnetic field orientation in Sect.~\ref{sec:homogeneity} and on the magnetic field strength in Sect.~\ref{sec:B_strength}.
We investigate the relationship between the magnetic field structure in the CMZ measured by {\PILOT} and recently proposed models for gas orbits in the Galactic Centre in Sect.~\ref{Sec:infinityloop}, using a model for the Milky Way's magnetic field that we describe in Appendix~\ref{sec:jbdmodel}. We discuss the comparison of our results with previous ground measurements in Sect.~\ref{sec:ground_measurements}. We summarize our conclusions in Sect.~\ref{sec:conclusions}. \\

\begin{figure*}[!ht]
	\centering
	\includegraphics[width=\textwidth]{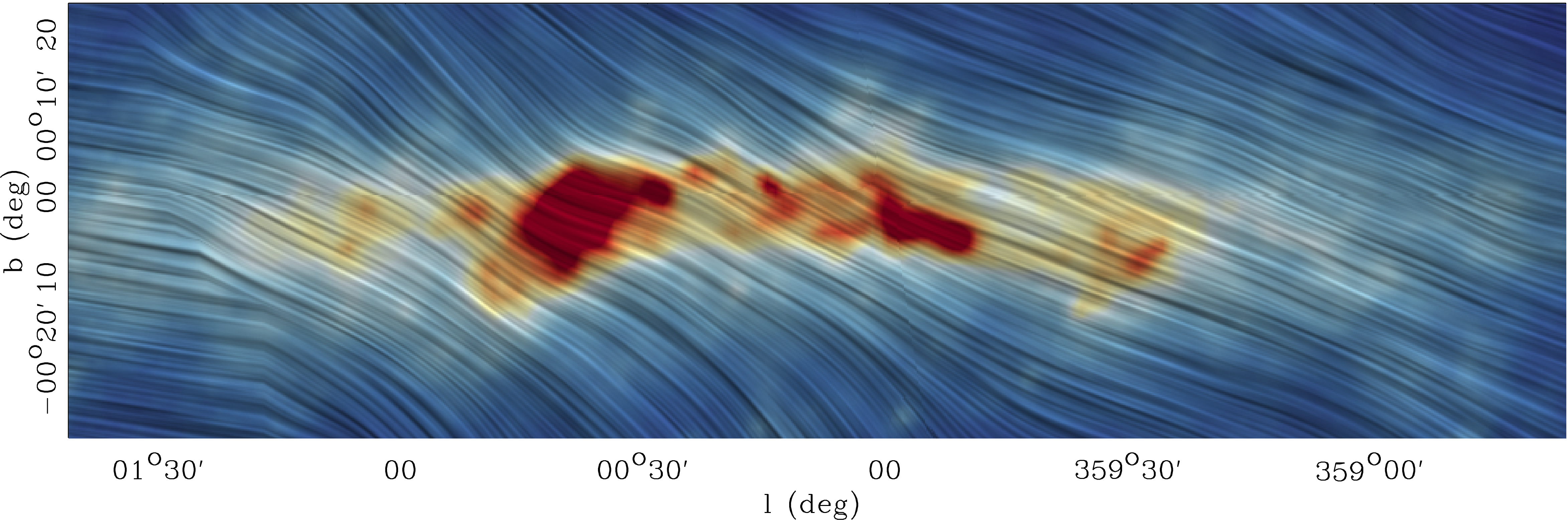}
	\includegraphics[width=\textwidth]{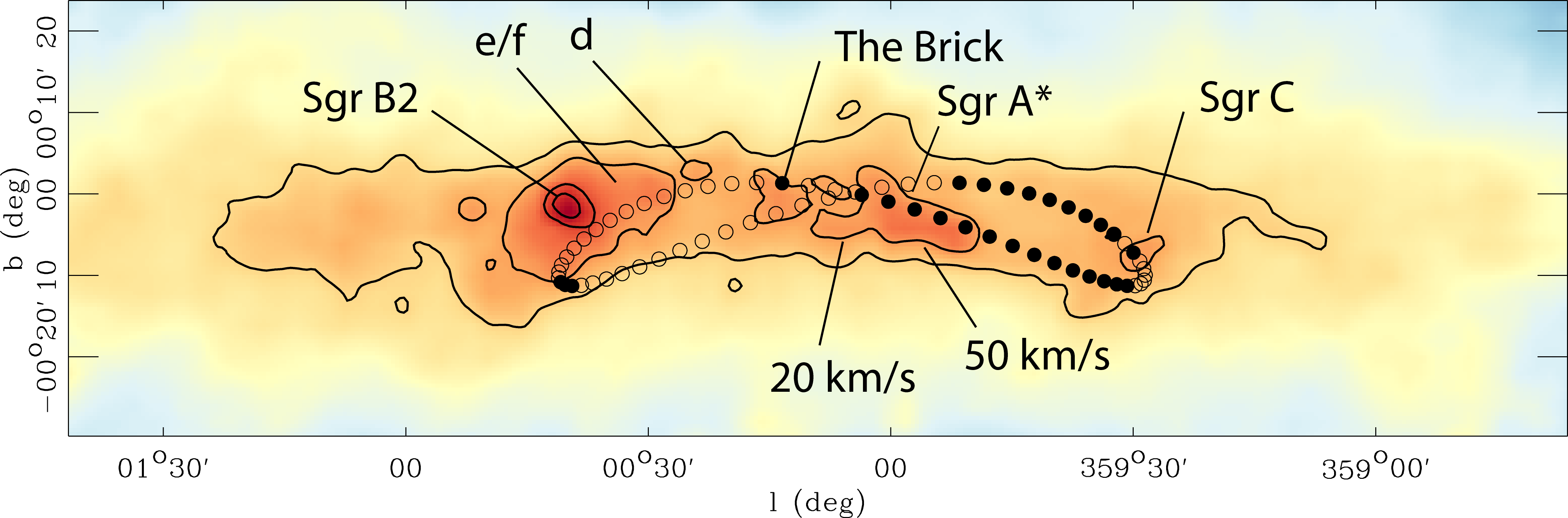}
	\caption{Upper panel: Map of the Galactic central molecular zone obtained with the {\PILOT} experiment at 240\,\micron. The color scale shows the total intensity in log scale. The overlaid texture based on the line integral convolution \citep{Cabral_Leedom1993} shows the orientation of the magnetic field projected on the plane, inferred from the measured dust polarization.
	Lower panel: Intensity as above, overlaid with total intensity contours at 0.1, 0.2, 0.5, and $1\times10^5$\,MJy/sr. The circles show the location of the 100-pc twisted ring as defined in \cite{Molinari_etal2011}. The filled symbols are discussed in Sect.~\ref{Sec:infinityloop}. The names of key molecular regions are overlaid.
	}
	\label{fig:schematic}
\end{figure*}

\noindent To aid discussion, a schematic diagram identifying key features of the CMZ that we refer to in this paper, overlaid on a map of the {\PILOT} $240$\,\micron\ intensity, is presented in Figure~\ref{fig:schematic}. Throughout the paper, we quote polarization angles in the GAL-COSMO convention (i.e. clockwise from Galactic North when looking at the source). We employ the acronym POS to describe quantities and structures that are projected onto the plane of the sky (e.g. the orientation of the magnetic field that is accessible via dust polarization measurements, which elsewhere in the literature is sometimes denoted as {$\Bperp$}), and the common acronym LOS to mean line-of-sight. 

\section{{\PILOT} Observations}
\label{sec:obs}

{\PILOT} observations of the GC region were obtained during the second flight of the {\Pilot} balloon experiment, which took place from Alice Springs, Australia, as part of the 2017 French Space Agency ({\CNES}) balloon campaign. A complete description of the {\Pilot} instrument is presented in \cite{Bernard_etal2016}, while the performance of the {\Pilot} instrument during the Alice Springs flight is described in \cite{Mangilli_etal2018}.

The data presented in this paper were obtained during four consecutive {\scenes} while the target field was at elevations between 16.5 and 56\,\degr. The total duration of the observations was $\simeq 32\,\textrm{min}$, including time for slewing and calibrations at the end of individual scans. The temperatures of the {\TRANS} and {\REFLEX} focal planes during the observations were stable, $304.96\pm{0.04}$\,mK and $309.07\pm{0.03}$\,mK respectively. The median balloon altitude was 39.8\,km, with peak-to-peak variation between 39.7 and 39.9\,km. The four {\scenes} used four distinct positions of the half wave-plate ({\HWP}) in order to uniformly sample polarization analysis directions as projected on the sky. Each of the {\scenes} was configured to scan the target region at a different angle, with a median scan speed of 11.8\,arcmin/s and a median scan leg duration of 13\,s.

These scan directions were chosen to be preferentially perpendicular to the Galactic plane, while at the same time providing sufficiently varied directions for efficient destriping during map-making (see Sect.\,\ref{Sec:Map_making}). The consequence of this observing strategy is that the elevation varies during scans, which causes variations of the signal due to emission from the residual atmosphere (see Sect.\,\ref{Sec:Atmospheric_signal_subtraction}). In Sect.\,\ref{Sec:Gain_calibration}, we describe how we use the elevation-dependent variations across the whole flight to obtain a focal plane map of the individual detector responses.

\section{Data processing}
\label{sec:Data_processing}

\subsection{Time constant correction}
\label{Sec:Time_constant_correction}

The procedure that we used to estimate the time constants of the bolometers is described in \cite{Mangilli_etal2018}. In summary, we use a combination of in-flight measurements of the internal calibration source ({\ICS}) and of strong glitches detected throughout the flight to derive the time constants of both the {\ICS} and of individual bolometers, which we assume to be constant throughout the flight. The timeline of each bolometer was then deconvolved from the corresponding time constant in Fourier space. The bolometer time constants are at most $\sim1$ sample ($=25$\,ms) and we estimate our knowledge of their value to be accurate within
$\simeq 2$ msec.

\subsection{Pointing reconstruction}
\label{sec:Pointing_reconstruction}

The pointing of each detector can be computed given the {\Estadius} stellar sensor information, the offset between the  {\Estadius} and {\PILOT} optical axes and a description of the {\PILOT} focal plane geometry.

The {\Estadius} stellar sensor \citep{Montel2015} data provides the pointing quaternion at a frequency of 4\,Hz 
with an accuracy of $\simeq 5$\,arcsec in both directions and $\simeq 15$\,arcsec in field rotation. During the observations of the CMZ region presented here, the performances of the {\Estadius} sensor were optimal.
The offset between the {\Estadius} and the {\PILOT} optical axes was found to vary during  flight, due to thermal and mechanical deformation of the instrument. This offset was monitored during the whole flight using bright sources. For that purpose, we used total intensity maps of individual {\scenes} of the {\PILOT} data obtained using the {\Scanamorphos} map-making algorithm (see Sect.~\ref{Sec:Map_making}). These maps were obtained using coordinates computed with preliminary {\Estadius} offsets derived during the flight from planet observations. These maps were correlated with {\Herschel} maps of the same sky region at 250\,\micron\ for different assumed values of the {\Estadius} offset. The best {\Estadius} offset for each {\scene} was derived as the one providing the best Pearson correlation coefficient between the {\PILOT} and {\Herschel} data. Note that three different bright sources were used in the {\Lzero} observations to derive {\Estadius} offsets. 

A preliminary version of the {\PILOT} focal plane geometry was obtained during ground calibration of the instrument \citep{Bernard_etal2016}. Here, we used all parameters of this determination including the pixel size and array rotation values. However, we refined the detector array position offsets with respect to the {\PILOT} focal plane center using in-flight measurements. For that purpose, we used a similar procedure as for deriving {\Estadius} offsets but applied to maps obtained with data from individual arrays. The offset difference between the individual offsets and the {\Estadius} offsets was adopted as the refined arrays offsets of the focal plane geometry.

The bolometer coordinates used here were derived combining the three quaternions describing the {\Estadius} pointing, the {\Estadius} offset with respect to the {\PILOT} focal plane center and each bolometer location in the focal plane. The {\Estadius} quaternions were interpolated in time to the time corresponding to each data sample, taking into account the time shift between individual detectors within one array caused by the time-multiplexing readout electronics. For a given {\scene}, the {\Estadius} offsets were assumed constant. The focal plane geometry was assumed invariant over all observations.

We estimate that the accuracy of the pointing from the differences between pointing reconstruction solutions obtained with different {\Estadius} offsets computed on the various sources used in {\Lzero}
is $\simeq15$\,arcsec.  

\subsection{Gain calibration}
\label{Sec:Gain_calibration}

\begin{figure*}[ht]
    \includegraphics[width=0.95\textwidth]{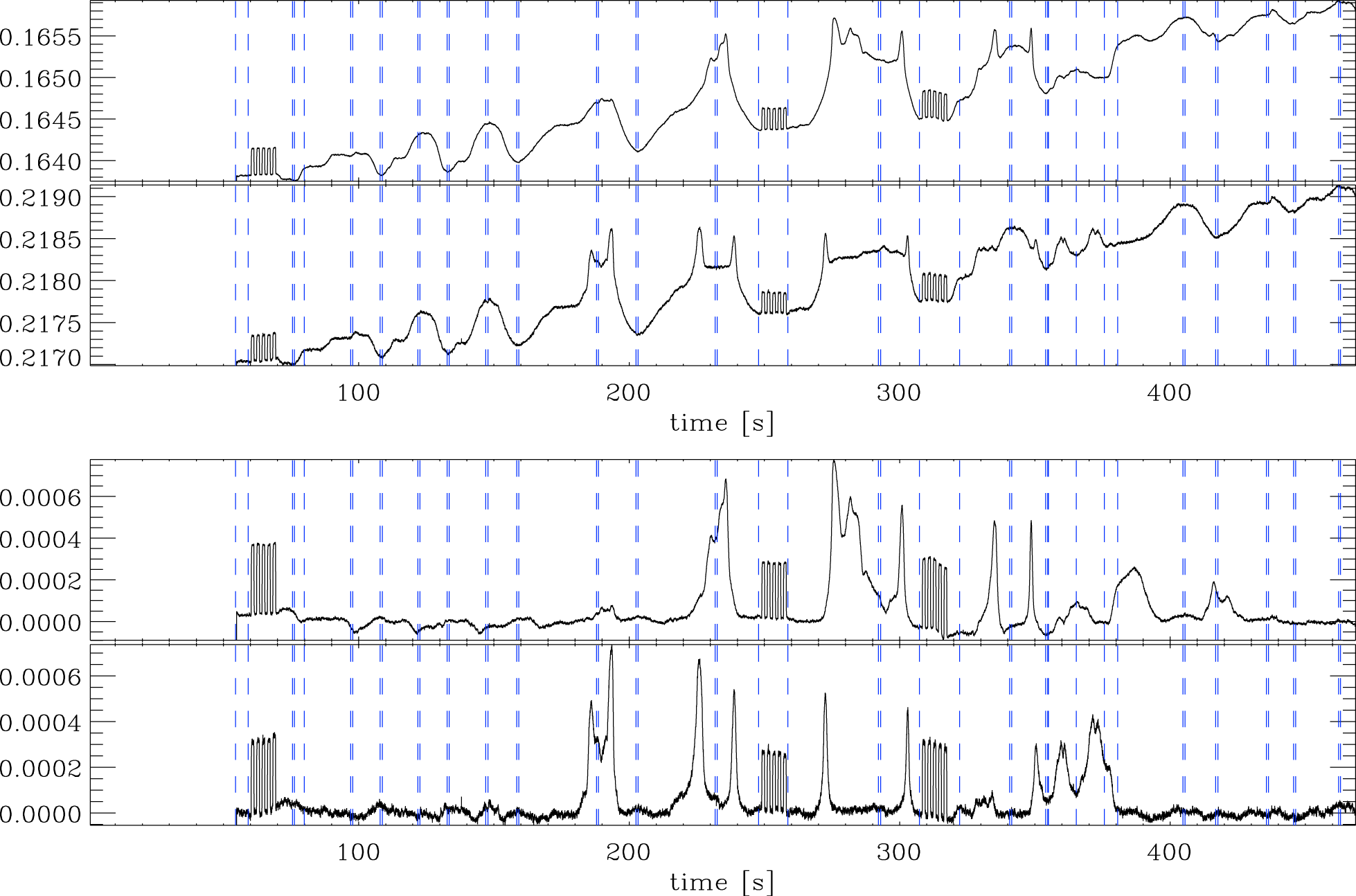}
	\caption{\label{fig:timelines} Calibrated {\PILOT} data for one {\scene} during the {\Lzero} observations. The upper panel shows the data before atmospheric signal removal. The lower panel shows the data with the atmospheric signal removed as described in Sect.\,\ref{Sec:Atmospheric_signal_subtraction}. In each panel, we show the array-averaged signal in $V$ as a function of time on the {\TRANS} array \#6 and and {\REFLEX} array \#7, from top to bottom. Vertical lines show the extent of individual scans. The modulated signal during calibration sequences on the {\ICS} at some ends of scans is clearly seen. 
	}
\end{figure*}

The gain calibration was performed as described in \cite{Mangilli_etal2018}. The raw data for each individual bolometer located by its position in the focal plane $(x,y)$, for each {\scene} (observed at a time $t_\textrm{tile}$), were divided by a response $\Resp(x,y,t_\textrm{tile})$, computed as:
\begin{equation}
\Resp(x,y,t_\textrm{tile})=\frac{R_\textrm{atmo} (x,y)}{\rho_\textrm{ICS} (x,y)}  \rho_\textrm{ICS} (x,y,t_\textrm{tile}),
\label{eq:resp}
\end{equation}
where $\Ratmo(x,y)$ is a normalized map of the relative response between all detectors, derived from the atmospheric signal averaged over the whole flight, $\rho_\textrm{ICS}$ is the {\ICS} signal averaged over the whole flight and $\rho_\textrm{ICS} (x,y,t_\textrm{tile})$ is the {\ICS} signal averaged over the {\scene}. $\Resp(x,y,t_\textrm{tile})$ is allowed to vary during the flight but is assumed to be constant within a given {\scene}. This gain calibration strategy assumes that the {\ICS} signal is intrinsically unpolarized.

From the all-flight time statistics on $\Resp$, we estimate that detectors gain inter-calibration is accurate at the percent level \cite{Mangilli_etal2018}. 
The absolute calibration of the data was obtained from the correlation between total intensity maps of {\Lzero} with {\Herschel} maps.

\subsection{Atmospheric signal subtraction}
\label{Sec:Atmospheric_signal_subtraction}

In order to isolate the astronomical signal from the residual atmospheric emission, we correlate the observed signal of each pixel with the pointing elevation consistent with the pointing solution described in Sect.~\ref{sec:Pointing_reconstruction}. The slope of the correlation is averaged over each {\scene}, and the correlated component of the signal is subtracted from the data. The top panel of Fig.~\ref{fig:timelines} shows the time evolution of the average signal measured on the five {\PILOT} arrays for one {\scene} before subtraction of the residual atmospheric emission for an {\scene} scanned with decreasing elevations. We can see that the timeline signal increases as the elevation decreases. In the bottom panel of Fig.~\ref{Sec:Atmospheric_signal_subtraction}, the atmospheric signal has been cleaned and the timeline shows a flat baseline.

\subsection{Map-making}
\label{Sec:Map_making}

\subsubsection{PILOT measurements}
\label{Sec:measurements}

The {\PILOT} signal can be expressed as described in \cite{Mangilli_etal2018} for a given bolometer at a position $(x,y) $ in the focal plane: 
\begin{equation}
m = R_{xy} T_{xy} \times [\StokesI \pm \StokesQ \cos(2\theta) \pm \StokesU \sin(2\theta)]+O_{xy},
\label{eq:pol_measure_easy_Stokes_sky}
\end{equation}
where $R_{xy}$ and $T_{xy}$ are the system response and optical transmission respectively, and $O_{xy}$ is an arbitrary electronics offset. For the configuration of the {\HWP} and analyzer in the {\PILOT} instrument, the $\pm$ sign is $+$ and $-$ for the {\REFLEX} and {\TRANS} arrays respectively (see \cite{Bernard_etal2016}).  $\theta=2\times\omega+\phi$ is the analysis angle. $\omega$ is the angle between the {\HWP} fast axis direction as projected on the sky and the horizontal direction measured counter-clockwise as seen from the instrument. In practice, the same patch of sky is observed at different times with at least two values of the analysis angle. $\phi$ is the time varying parallactic angle measured counterclockwise from equatorial North to Zenith for the time and direction of the current observation. The precise positioning of the {\HWP} fast axis was measured from ground calibration measurements obtained in front of a polarized signal of known polarization direction.
From the variations between various calibration sequences, we estimate that $\omega$ is known to an accuracy of $\simeq 1\degr$.

The light polarization fraction $\polfrac$ and polarization orientation angle $\polang$ are then defined as:
\begin{equation}
\label{equ:pem}
\polfrac=\frac{\sqrt{\StokesQ^2+\StokesU^2}}{\StokesI}
\end{equation}
and
\begin{equation}
\label{eq:thetam}
\polang=0.5 \times \arctan(\StokesU/\StokesQ).
\end{equation}
The magnetic field orientation is defined by $\polang$ rotated by 90$^\circ$.
Note that the $\StokesQ$ and $\StokesU$ in Eq.\,\ref{eq:pol_measure_easy_Stokes_sky}, and therefore $\polang$ are in the {\IAU} convention, i.e., positive counterclockwise on the sky, and measured with respect to Equatorial North (corresponding to $\polang=0\,\deg$). Comparison to data taken in different convention, such as the {\Planck} data, which are delivered in the {\COSMO} convention with respect to Galactic North, require changing the angle convention (which can be done simply as $\StokesQ_{C} = \StokesQ_{I}$) and $\StokesU_{C} = -1 \times \StokesU_{I}$) and the appropriate coordinate rotation of $\StokesQ$ and $\StokesU$.
In this paper, since we study the orientation of the magnetic field in the Galactic plane, we discuss values of the polarization angles in Galactic coordinates and in the {\COSMO} convention (as defined in the {\sc HEALPix} software packages \citep{healpix}).
In this paper, we also refer to the {\Sfunction} $\DeltaAng$ which quantifies the regularity of the $\Bfield$ orientation measured defined as: \citep[see]{Planck2015_intermediate_XIX} as
\begin{equation}
\DeltaAng^2 (\vec{x},\lag)=\frac{1}{N} \sum_{i=1}^{N}{ 
(\polang(\vec{x})-\polang(\vec{x}+\vec{\lag}_i))^2
},
\label{equ:delta_phi1}
\end{equation}
where $\vec{x}$ represents the central pixel, and $\polang(\vec{x}+\vec{\lag}_i)$ the polarization angle at a sky position displaced from the centre by the displacement vector $\vec{\lag}_i$. The average in Eq.\,\ref{equ:delta_phi1} is taken over an annulus of radius $\lag=|\vec{\lag}|$ (the ``lag'') and width $\Delta\lag$ around the central pixel and containing $N$ pixels. $\DeltaAng$ can be computed at varying angular resolution of the data and with a lag usually ranging from half the data FWHM to arbitrarily large scale, and $\DeltaAng$ increases with $\lag$ as the field structure decorelates with increasing scale.

Inversion of the observed signals to derive sky maps of $\StokesI$, $\StokesQ$ and $\StokesU$ was done through polarization map-making algorithms. We used two different map-making algorithms,  {\scanamorphos} and {\ROMA}, which are described in Sect.\,\ref{Sec:scanamorphos} and \ref{Sec:Roma} respectively. We checked that the two algorithms give consistent results, and used the {\scanamorphos} maps for the data analysis, which appeared of better quality at the current stage of the processing.

\subsubsection{{\scanamorphos}}
\label{Sec:scanamorphos}

\begin{figure*}[ht]
\centering
\includegraphics[width=0.31\textwidth]{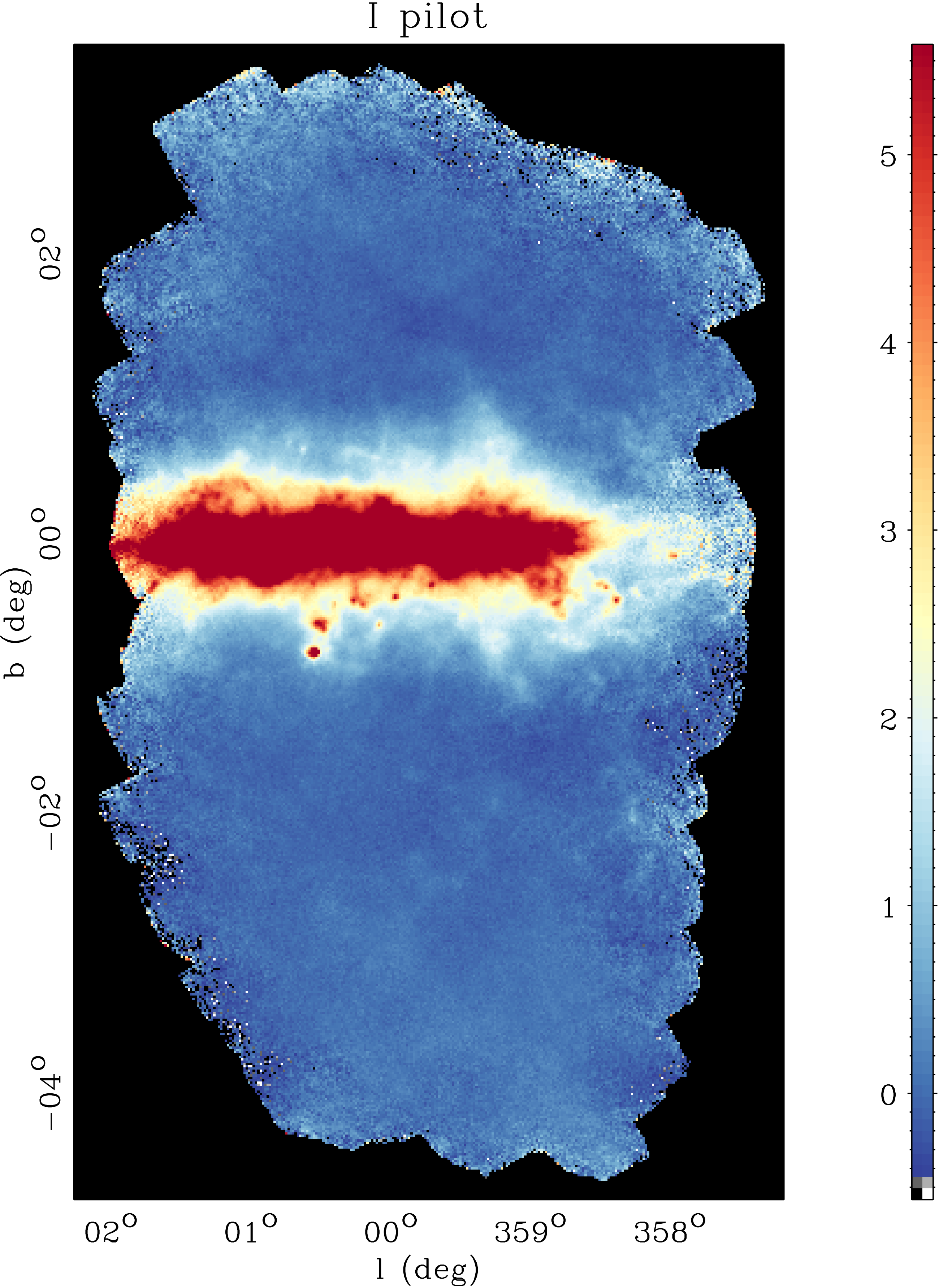}
\includegraphics[width=0.31\textwidth]{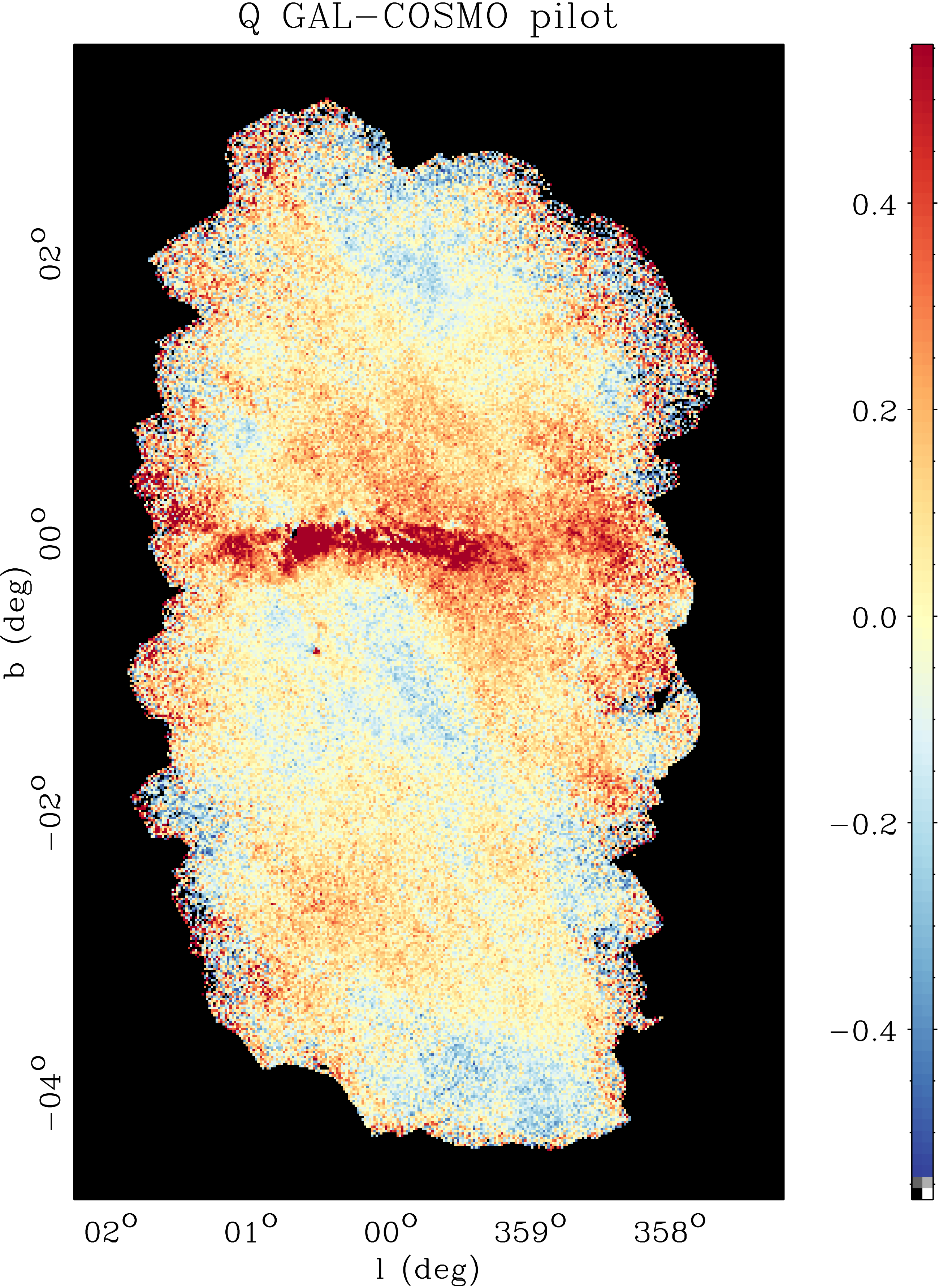}
\includegraphics[width=0.31\textwidth]{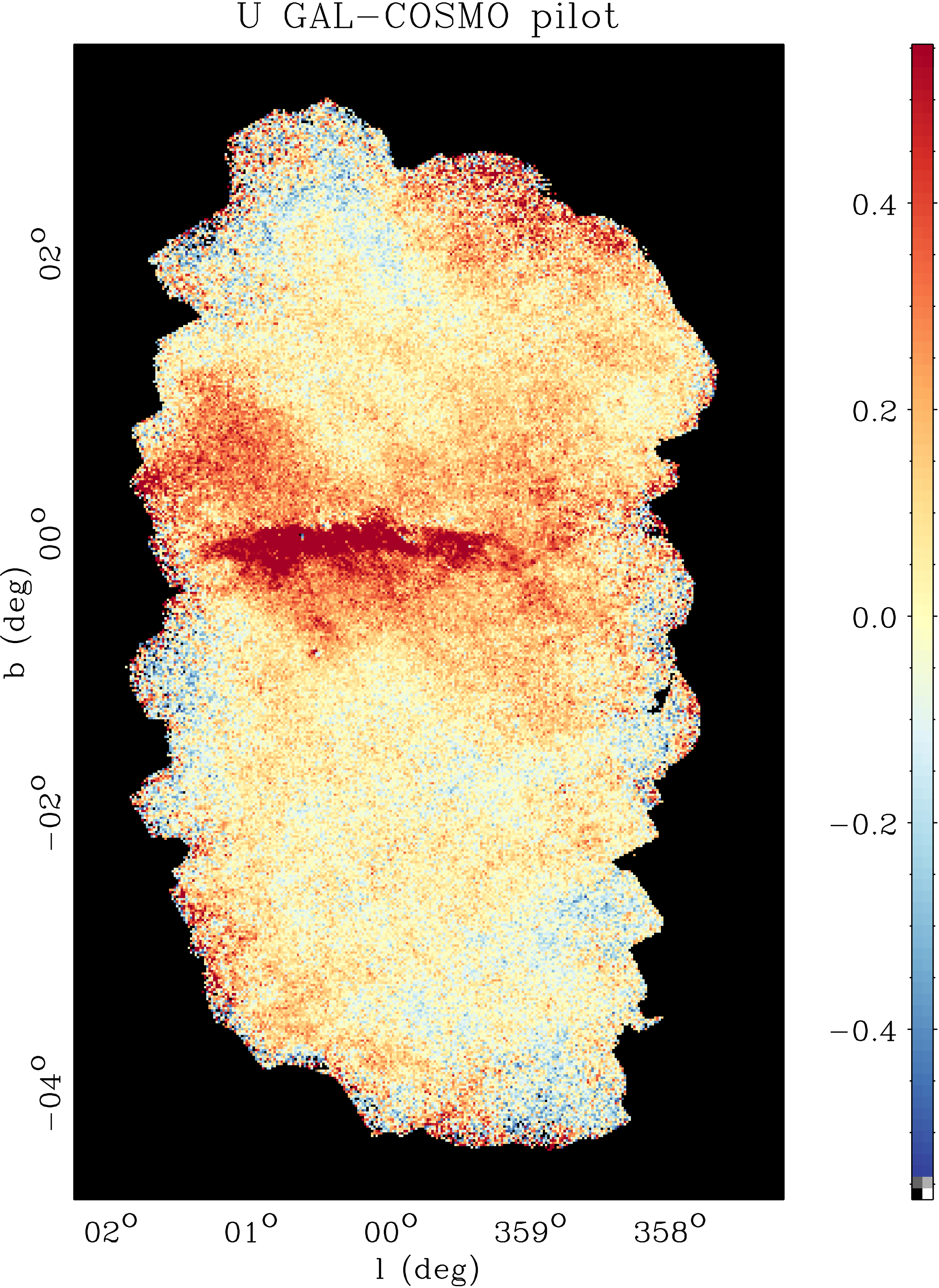}
\caption{\label{fig:IQU_scanam_largescale} {\PILOT} maps of the $\StokesI$, $\StokesQ$, $\StokesU$ parameters obtained towards the galactic center region, processed using the {\scanamorphos} software. The maps are shown at the full resolution of the instrument of 2.2\,arcmin. The maps are in units of $10^3$\,MJy/sr.}
\end{figure*}

\begin{figure*}[ht]
\centering
\includegraphics[width=0.91\textwidth]{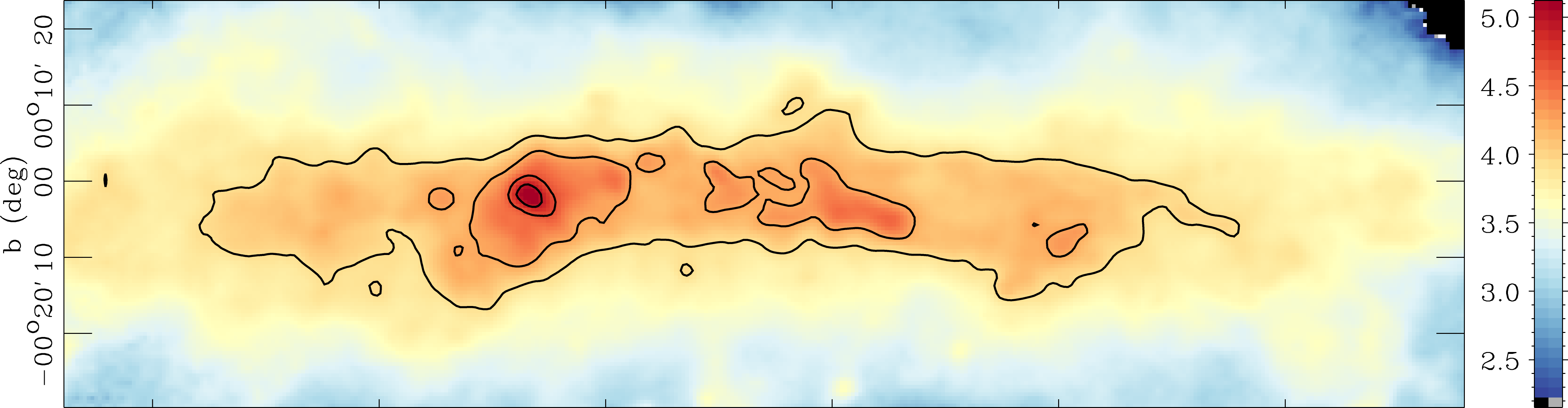}
\includegraphics[width=0.91\textwidth]{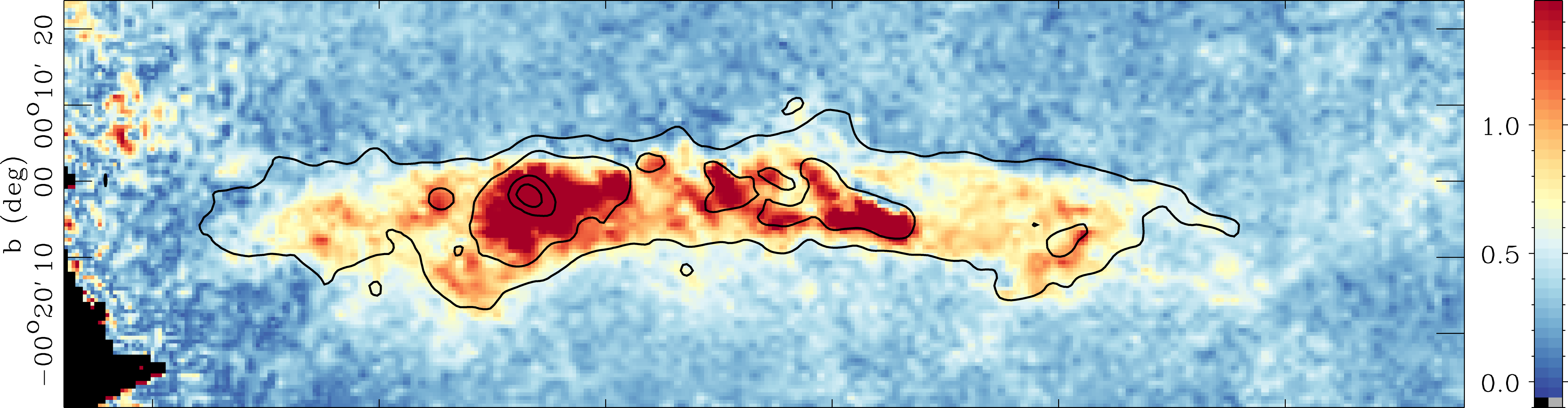}
\includegraphics[width=0.91\textwidth]{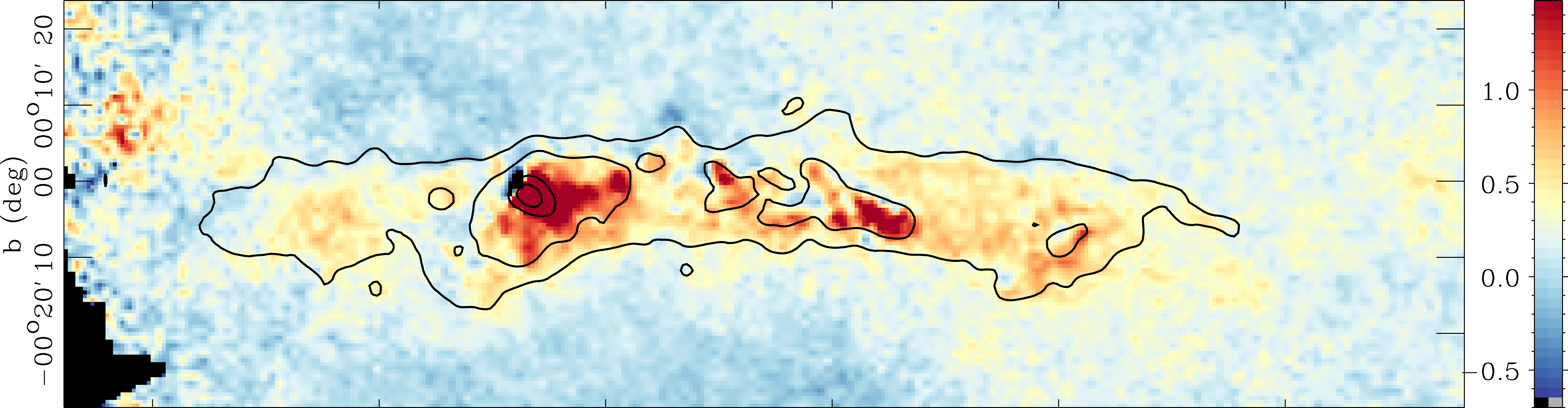}
\includegraphics[width=0.91\textwidth]{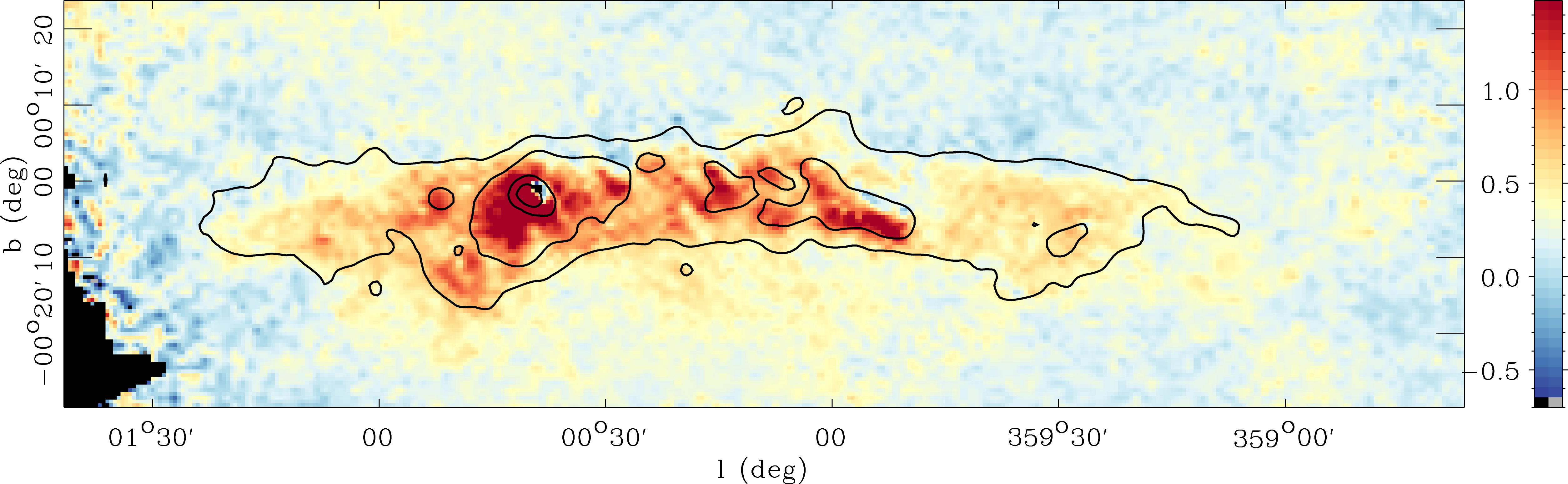}
\caption{\label{fig:IQU_scanam_zoomed} {\PILOT} maps of, from top to bottom, the  total intensity $\StokesI$, the polarized intensity $P$, the Stokes $\StokesQ$ and $\StokesU$ parameters, zoomed on the CMZ region and processed using the {\scanamorphos} software. The maps are shown at the full resolution of the instrument of 2.2\,arcmin. The intensity map is shown in log scale. The maps of $P$, $\StokesQ$ and $\StokesU$ are in units of $10^3$\,MJy/sr.
}
\end{figure*}

\begin{figure*}[ht]
\includegraphics[width=1.0\textwidth]{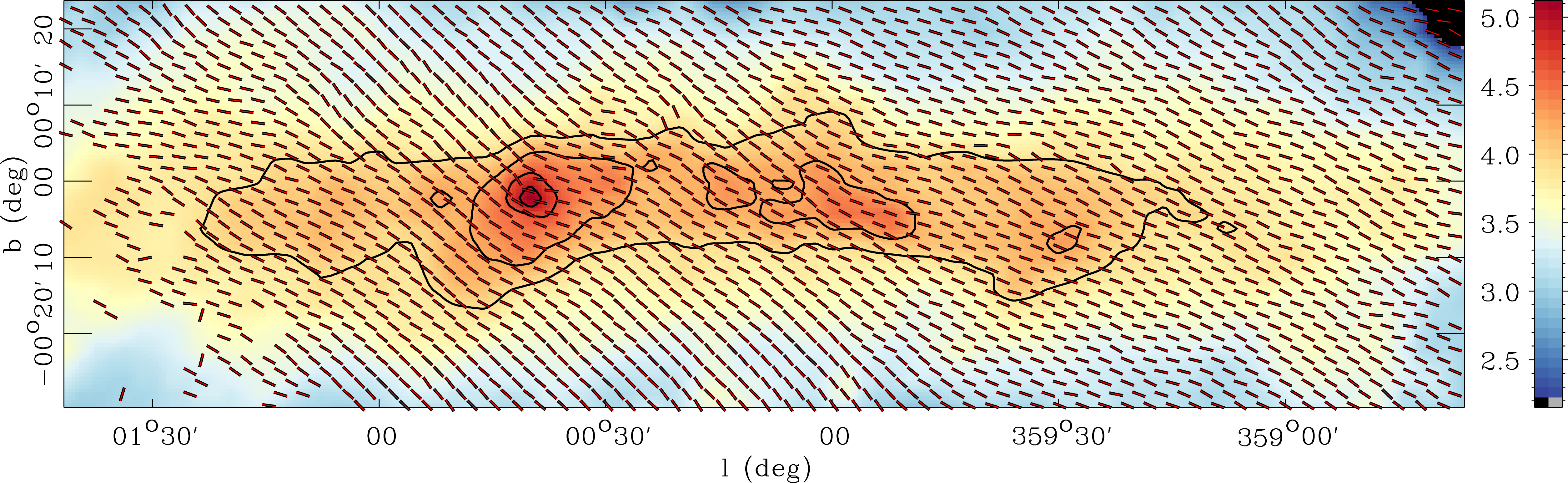}
\includegraphics[width=1.0\textwidth]{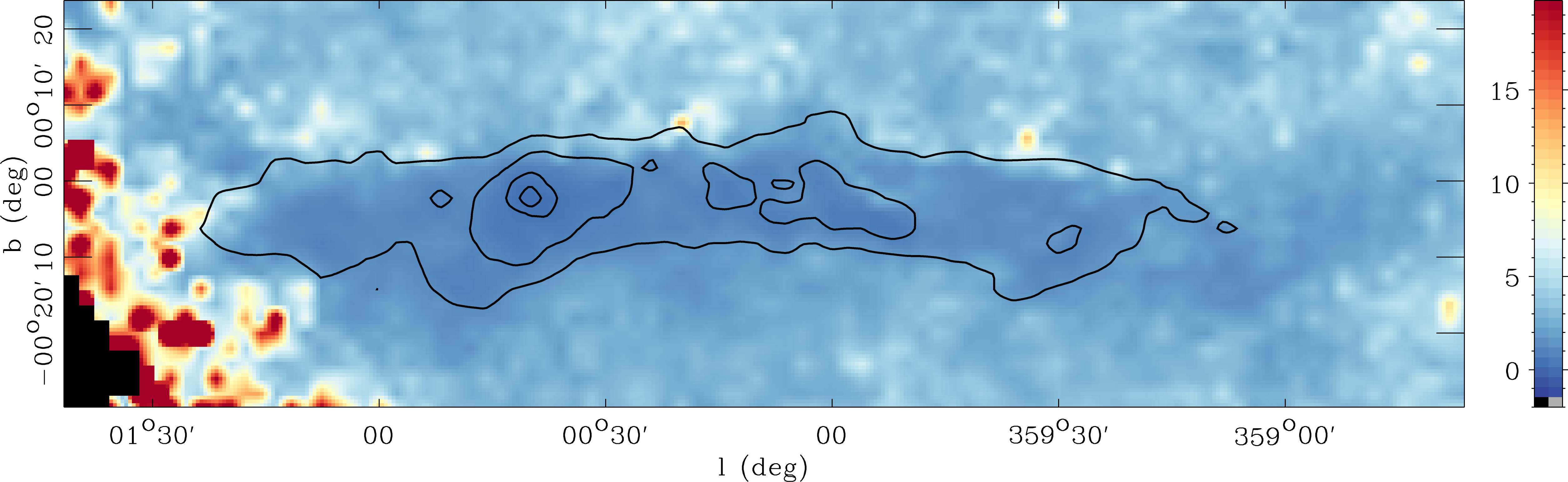}
\caption{\label{fig:polar_vector_scanam} Upper panel: Map of the $B$-field orientation as derived from the {\PILOT} data using the {\scanamorphos} software. The total intensity is shown as the color map, in log scale. The $B$-field orientation is shown as constant length segments orthogonal to the polarization orientation, where $\sigma_\psi < 8\,\degr$.
Lower panel: Map of the polarization angle uncertainty per beam, in degrees.
}
\end{figure*}

The {\Scanamorphos} software was initially developed to process on-the-fly data from the PACS and SPIRE bolometer arrays on-board {\Herschel} \citep{Roussel2013}. The software has subsequently been refined to cope with the dominant atmospheric emission in ground-based observations for different instrument architectures, such as {\Artemis} on {\APEX} \citep{Roussel2018} and {\NIKAdeux} on the IRAM 30-m telescope. It has also been adapted to process the polarization data of {\PILOT} as outlined below. Its core principle is to exploit all the available redundancy in the observations, assuming only that the sky brightness is invariant in time and that the component of the low-frequency noise that is uncorrelated (termed individual drifts or flicker noise in the documents describing the code) has a symmetric probability distribution around zero, without necessarily being Gaussian. The full processing comprises the following tasks, most of them iterative: 1) slicing of scans into distinct scan legs, allowing for a large variability of the scanning speed inherent to balloon observations; 2) baseline subtraction using robust linear fits to the time series of the whole observing tiles, blocks of 4 scan legs, and individual scan legs, over a source mask that is automatically built and updated; 3) destriping using baseline subtraction in which the fits are constrained by the data redundancy, in order to provide a coherent solution for all scan legs and all detectors; 4) masking of glitches detected exploiting the redundancy to avoid masking true sources; 5) subtraction of the average drift on small timescales (i.e., the common mode of both the residual low-frequency noise and the residual atmospheric signal); 6) subtraction of the individual drifts on successively smaller timescales; 7) projection of the corrected data. The definition of the time and space grids used by these tasks and the algorithm are described in the companion paper of the code developed for {\Herschel}.

In order to deal with polarization data, ideally three time series corresponding to the $\StokesI$, $\StokesQ$ and $\StokesU$ Stokes parameters should be considered in the modules subtracting the drifts, instead of a single brightness time series. Even in the absence of noise, the recorded signal at a given location varies as a function of time and detector, due changes in the {\HWP} position, focal plane position and parallactic angle as given in Eq.\,\ref{eq:pol_measure_easy_Stokes_sky}. However, the noise-free $\StokesI$, $\StokesQ$ and $\StokesU$ series should be invariant. The Stokes parameters are recomputed at each iterative step following the formalism of Eq.\,\ref{eq:pol_measure_easy_Stokes_sky}. The measured signal at a given time and for a given detector can then in principle be compared to the signal predicted from $\StokesI$, $\StokesQ$ and $\StokesU$ series computed from all the available data. But the signal depends on $\StokesQ$ and $\StokesU$ through sinusoidal terms, whose argument depends on time and detector. To avoid modulating the errors on $\StokesQ$ and $\StokesU$ when computing signal averages or differences, we proceed in a slightly different way, by subtracting the simulated $\StokesQ$ and $\StokesU$ terms from the measured signal to generate a "noised intensity" time series, that can be directly compared to the simulated intensity. In case where there is not enough redundancy in the analysis angles for a given pixel, i.e. when these are distributed over fewer than 4 separate groups, $\StokesQ$ and $\StokesU$ are not computed and the total intensity is equated to twice the average signal. This occurs only on the edges of the maps. In practice, the initial data are too noisy to allow meaningful estimates of the Stokes parameters, leading us to perform the first few drift subtraction steps as if the data were not polarized (i.e. assuming $\StokesQ = \StokesU = 0$). It is only when the major part of the average drift has been subtracted that we take polarization into account.

For the observations discussed here, the data was taken in a way such that three observing tiles out of four were taken with roughly the same scan direction, the fourth tile being taken roughly in the orthogonal direction. We have explored the possibility of modifying the reference $\StokesI$, $\StokesQ$, $\StokesU$ maps in the {\scanamorphos} destriping module. These are normally computed from the whole  data at the previous iteration of the destriping, but it is possible to lower the weight of some tiles, or to provide independent maps from simulations
, for example. We have not used this option for the present work.
Given the median scan speed and the sampling rate of the data, the fiducial value of the noise stability length is 1.5 times the beam {\FWHM}, and we have set it to twice the {\FWHM} to increase sample statistics.

The timelines described in Sect.\,\ref{Sec:Atmospheric_signal_subtraction} were provided to the code, without glitch masking, since glitches are identified internally to {\scanamorphos}. We provided timelines with the atmospheric contribution removed, but checked that maps computed without this initial correction are compatible with the ones presented here, thanks to the ability of {\scanamorphos} to subtract common modes.
The resulting $\StokesI$, $\StokesQ$, $\StokesU$ maps obtained over the extent of the whole region mapped are shown in Fig.\,\ref{fig:IQU_scanam_largescale}.
The structure of the magnetic field inferred is shown in Fig.\,\ref{fig:polar_vector_scanam}. The uncertainty on the polarization angle are also shown in Fig.\,\ref{fig:polar_vector_scanam}.
These maps are discussed in Sect.\,\ref{Sec:Polarization_results}.

\subsubsection{{\ROMA}}
\label{Sec:Roma}

The {\ROMA} code, described in \cite{deGasperis+2016}, 
was developed for cosmology experiments aiming at measuring the polarization of the Cosmic Microwave Background (CMB). For this reason it is optimized for noise dominated timelines.
The code assumes stationnarity of the noise statistical properties, which is reasonably the case for the data of {\PILOT} flight\#2 \citep[see][]{Mangilli_etal2018}.
{Under the assumption of a Gaussian and stationary noise, the {\ROMA} code adopts a Fourier-based, preconditioned conjugate gradient iterative method to solve for the maximum-likelihood signal map yield by the Generalized Least Square (GLS) approach \citep[see][for details]{deGasperis+2016}.

In particular, given the pointing matrix $\mathbf{A}$ and the timeline vector $\mathbf{D}$, the optimal GLS solution for the signal estimate $\mathbf{\widetilde S}=(\mathbf{\widetilde I}, \mathbf{\widetilde Q}, \mathbf{\widetilde U})$ is:

\begin{equation}
 \mathbf{\widetilde S} = \left( \mathbf{A}^T
\mathbf{N}^{-1} \mathbf{A}\right)^{-1}
  \mathbf{A}^T \mathbf{N}^{-1} \mathbf{D},
\label{eq:RomaGLS}\end{equation}

where $\textbf{N} \equiv \left\langle \textbf{n}_t^{\ } \textbf{n}_t' \right\rangle$ is the noise covariance matrix in the time domain ($\textbf{n}_t$ indicates the instrumental noise at time $t$). Notice that the matrix $\textbf{N}$ is block diagonal only in case of no noise cross-correlation.} 

The {\PILOT} timelines of the four galactic center region observations are provided to the code after gap-filling of calibration sequences at the ends of scans using a white noise realization with the white noise level of each detector. The glitches are also masked and the atmospheric contribution is removed as described in \ref{Sec:Atmospheric_signal_subtraction}.

In order to ensure that the signal is noise dominated, we subtracted from the observed timeline a simulated timeline based on the {\Planck} 353 GHz total intensity map {extrapolated to the {\PILOT} frequency}.
The subtracted component is assumed unpolarized and is therefore equal on the {\TRANS} and {\REFLEX} detectors apart for possible pointing mismatch. The difference timeline is processed through {\ROMA} with 100 iterations that are enough to ensure a good convergence. The map used for the signal subtraction is added back to the resulting total intensity map.
The noise covariance matrix provided to the {\ROMA} code was determined from the {\PILOT} observations of the {\BICEP} field obtained during the same flight \citep{Mangilli_etal2018}, treated in the same way as the {\Lzero} data. The map resolution is set to $N_\mathrm{side}=2048$ (1.7\,arcmin).
The code also provides maps of the variances on the Stokes parameters as well as a map of the inverse conditioning matrix allowing to assess if a sufficient number of data taken at different {\HWP} angles are available for a given pixel. The pixel inverse condition number (ICN) is defined as the ratio of the absolute values of the smallest and largest eigenvalue of the preconditioning matrix, and provides a good tracer of the errors in the map due to a non ideal angle coverage on the given pixel (a value of $\textrm{ICN}=0.5$ means uniform angle coverage). In our case, pixels with $\textrm{ICN}\le 10^{-2}$ are removed from the analysis.

\begin{figure*}
    \centering
    \includegraphics[width=0.3\textwidth]{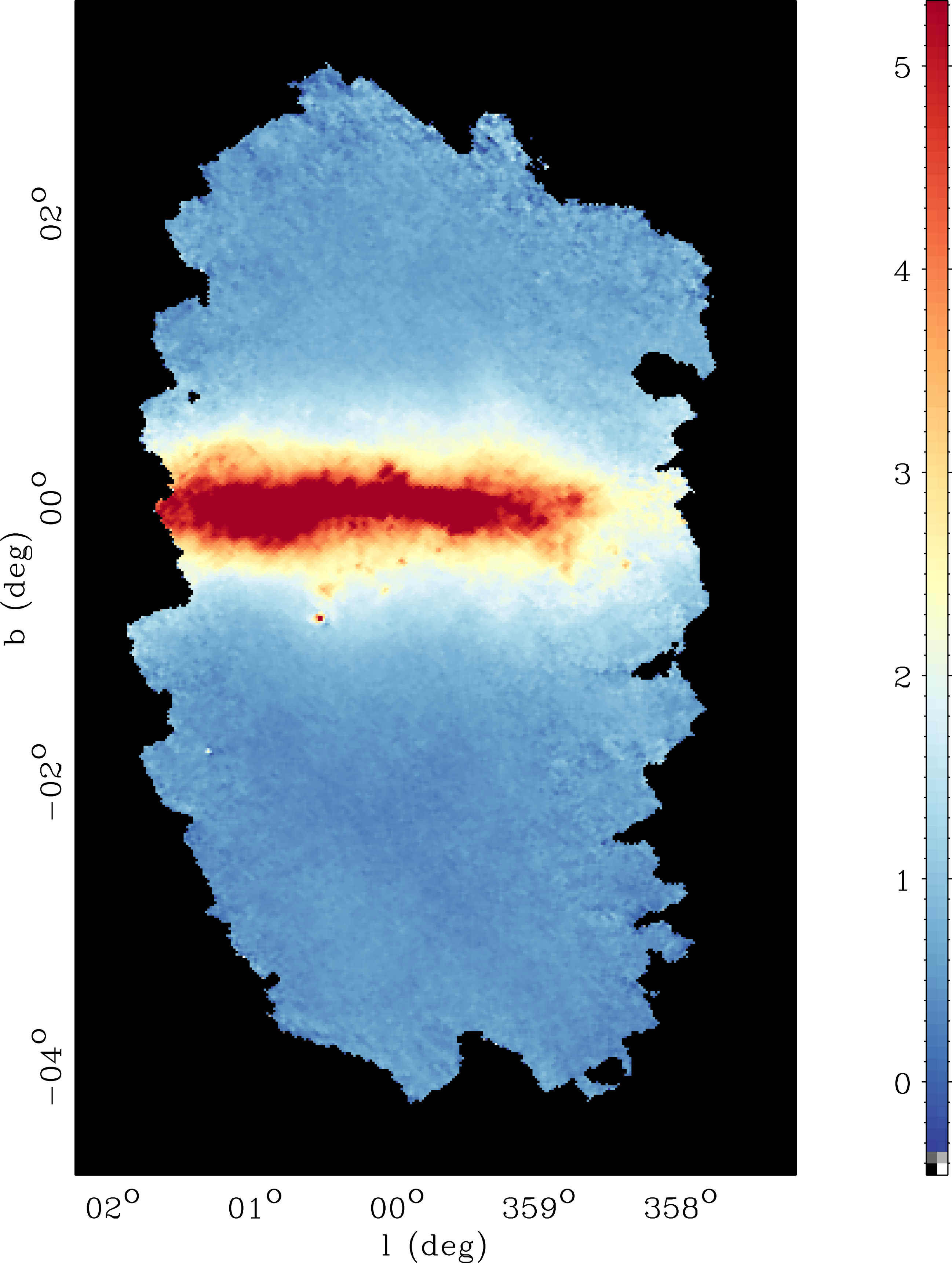}
    \includegraphics[width=0.3\textwidth]{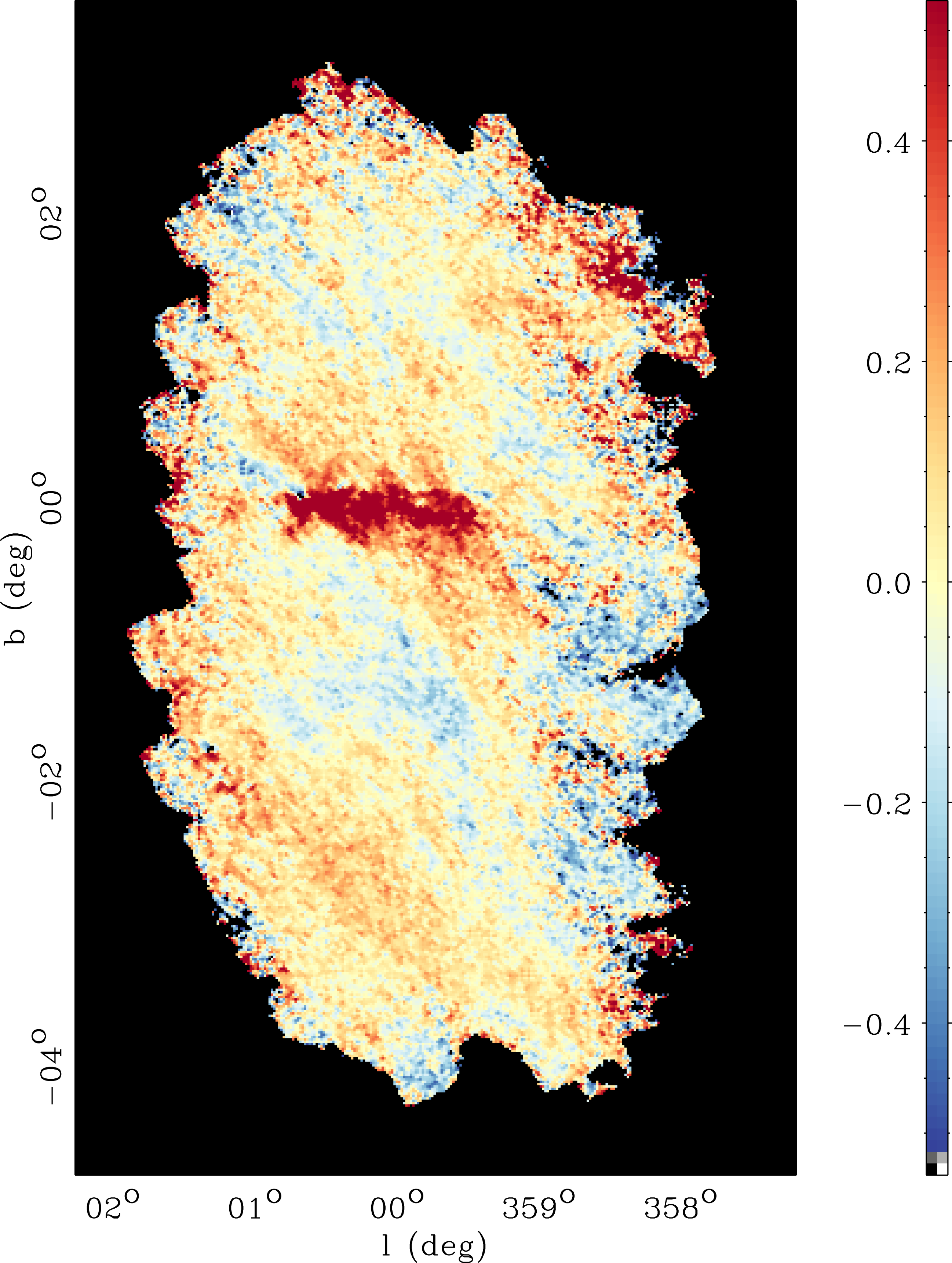}
    \includegraphics[width=0.3\textwidth]{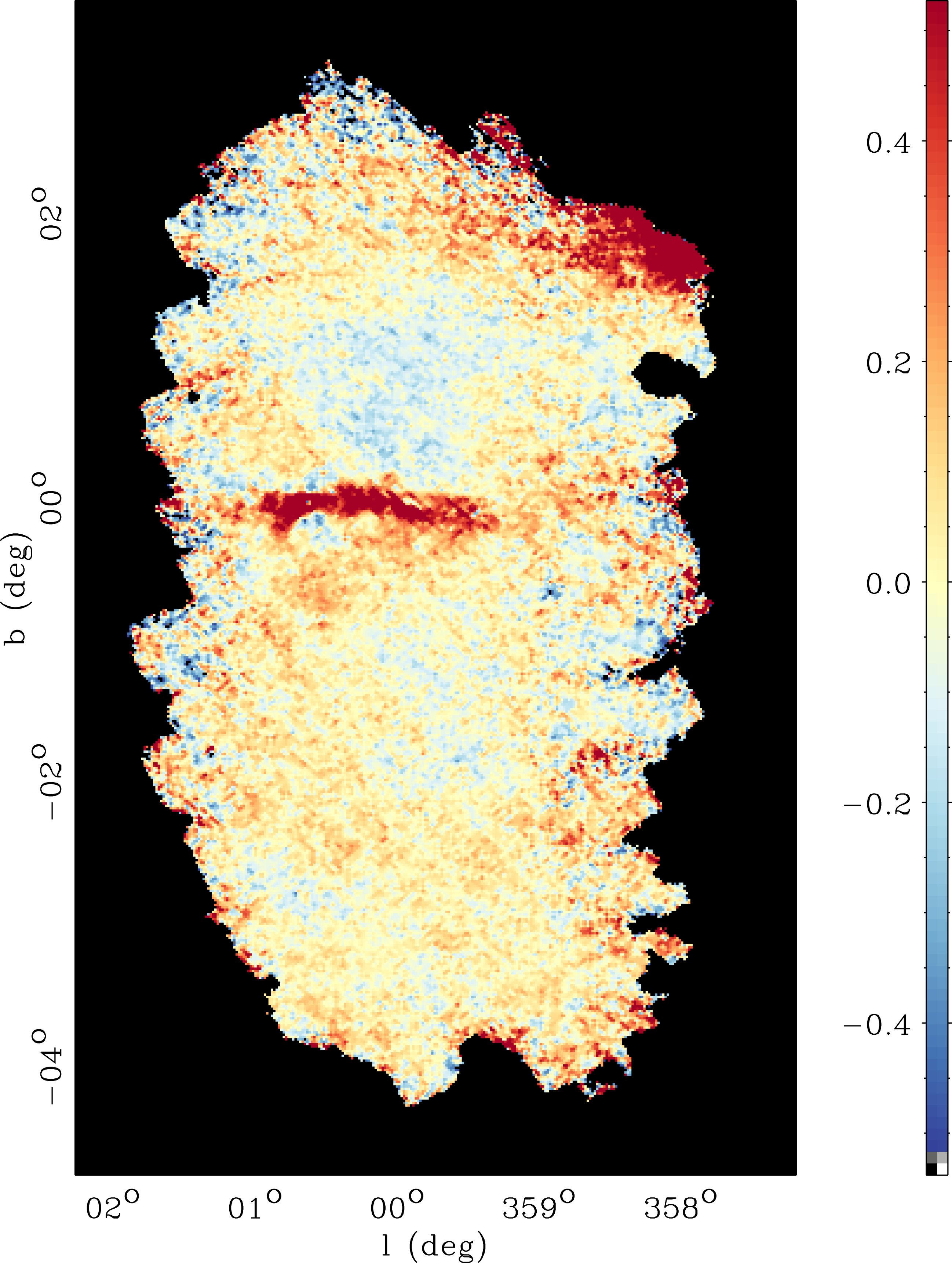}
    \caption{$PILOT$ large scale maps of the $I$, $Q$ and $U$ Stokes parameters obtained with the {\ROMA} software. The maps are in units of $10^3$\,MJy/sr.
    }
    \label{fig:ROMA_IQU}
\end{figure*}

The $\StokesI$, $\StokesQ$, $\StokesU$ maps obtained are shown in Fig.\,\ref{fig:ROMA_IQU}. Due to the difficulty of accurately modelling the noise covariance matrix, in particular the off-diagonal terms that are not included in the analysis, the quality of the {\ROMA} polarization maps is degraded and residuals stripes are still present, in particular in the $Q$ map. 
For this reason we will not use the {\ROMA} maps as the baseline for the analysis. Still, the {\ROMA} maps allow for validation tests of the baseline results, as discussed in Sect. \ref{Sec:mapmaking_comparison}

\section{Polarization results}
\label{Sec:Polarization_results}

\subsection{Polarization maps}
\label{sec:pilot_maps}

\noindent The final maps of Stokes $\StokesI$, $\StokesQ$ and $\StokesU$ parameters across the full Galactic Center field observed by {\PILOT} are shown in Figure~\ref{fig:IQU_scanam_largescale}. In this paper, we focus on the polarization results for the $\sim3.1\,\degr \times 0.9\,\degr$ CMZ region, for which we show zoomed maps of $\StokesI$, $\StokesQ$ and $\StokesU$ in Figure~\ref{fig:IQU_scanam_zoomed}. The corresponding map of the polarization angle $\psi$, calculated according to Equation~\ref{eq:thetam} is shown as an overlay in Figure~\ref{fig:polar_vector_scanam}. 

\noindent A common issue for the interpretation of polarization observations is how the observed polarization angles (and inferred magnetic field structure) projected onto the plane of the sky (POS) correspond to emission regions situated along the line-of-sight. This is especially of concern for observations in the Galactic Plane. In Appendix~\ref{sec:gal_model_annex}, we present a simple model that we have developed to quantify the line-of-sight (LOS) depth through the Galactic Plane that is probed by our observations. From this model, we conclude that the CMZ makes the dominant contribution to our observed polarisation signal towards the Galactic Center. This is qualitatively confirmed by Figure~\ref{fig:IQU_scanam_zoomed}, where it is evident that the structure in the {\PILOT} $\StokesQ$ and $\StokesU$ maps (as well as $\StokesI$) is dominated by well-known emission features of the CMZ, e.g., Sgr~B2, the Brick and the 20\,\kms\ cloud (Fig~\ref{fig:schematic}) \citep[e.g.][ and references therein]{kauffmann_etal17}.

\begin{figure}[!ht]
\centering
\includegraphics[width=0.5\textwidth]{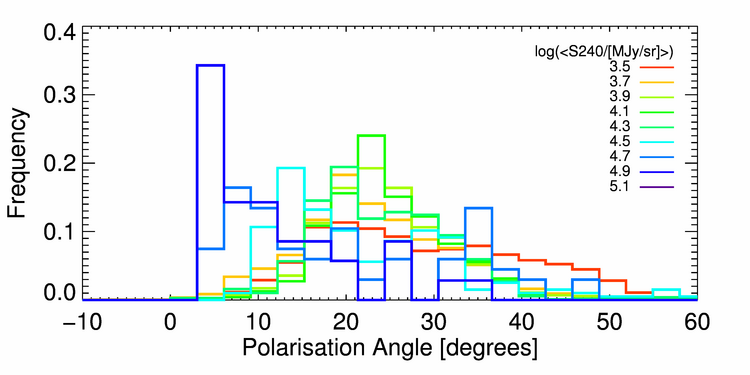}
\includegraphics[width=0.5\textwidth]{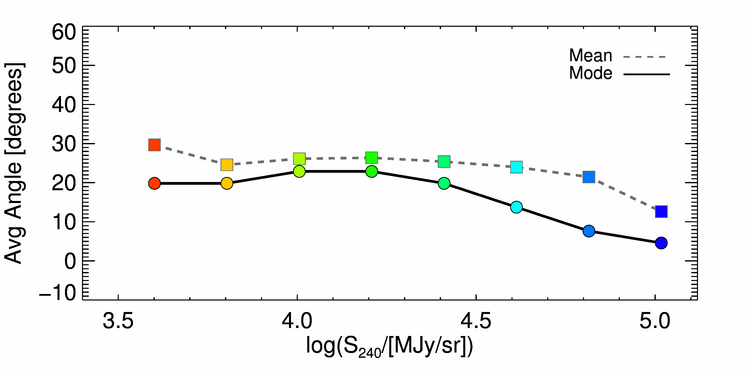}
\caption{Dust polarization angles measured in the CMZ as a function of 240\,\micron\ intensity. The top panel shows the polarization angle distributions plotted separately for intervals of increasing 240\,\micron\ intensity in the {\PILOT} \StokesI map. The characteristic 240\,\micron\ brightness is indicated using colour, and ranges from $10^{3.5}$\,MJy/sr (red) to $10^{5.1}$\,MJy/sr (blue).  The bottom panel plots the mean and mode of the polarization angle distributions directly as a function of the 240\,\micron\ intensity.}
\label{fig:l0_meanpsi_s240}
\end{figure}

\noindent What is most striking about the {\PILOT} polarization data for the CMZ is that the polarization angle is relatively uniform across the entire region shown in Fig.\,\ref{fig:polar_vector_scanam}, with a dominant orientation of $\sim+22\,\degr$. This corresponds to a dominant POS magnetic field orientation of $\sim+22\,\degr$ with respect to the Galactic Plane. Close inspection of the polarization vectors in Fig.\,\ref{fig:polar_vector_scanam} suggests that the polarization angle diverges from this overall mean orientation in regions of high brightness e.g., near the Sgr~B2, 20\,\kms\ and 50\,\kms\ clouds. We quantify this in Fig.\,\ref{fig:l0_meanpsi_s240}, where we show the distribution of polarization angles within regions of increasing 240\,\micron\ intensity in the $\StokesI$ map. Polarization angle measurements with high uncertainty ($\sigma (\psi) > 8\,\degr)$ are excluded from the analysis. The peak of the polarization angle distribution measured for high brightness regions (shown as blue histograms in the top panel) is clearly shifted towards lower values compared to the polarization angles measured in regions of lower intensity. 
We discuss possible explanation for this trend in Sect.~\ref{sec:homogeneity}, where we make a preliminary estimate for the average magnetic field strength in the CMZ based on the POS field structure that we infer from the PILOT dust polarization measurements.\\

\subsection{Comparison between map-making methods}
\label{Sec:mapmaking_comparison}

We take advantage of having two independent map-making methods to compare the polarization angles results obtained with the two pipelines.

\begin{figure}[!t]
    \centering
\includegraphics[width=0.5\textwidth]{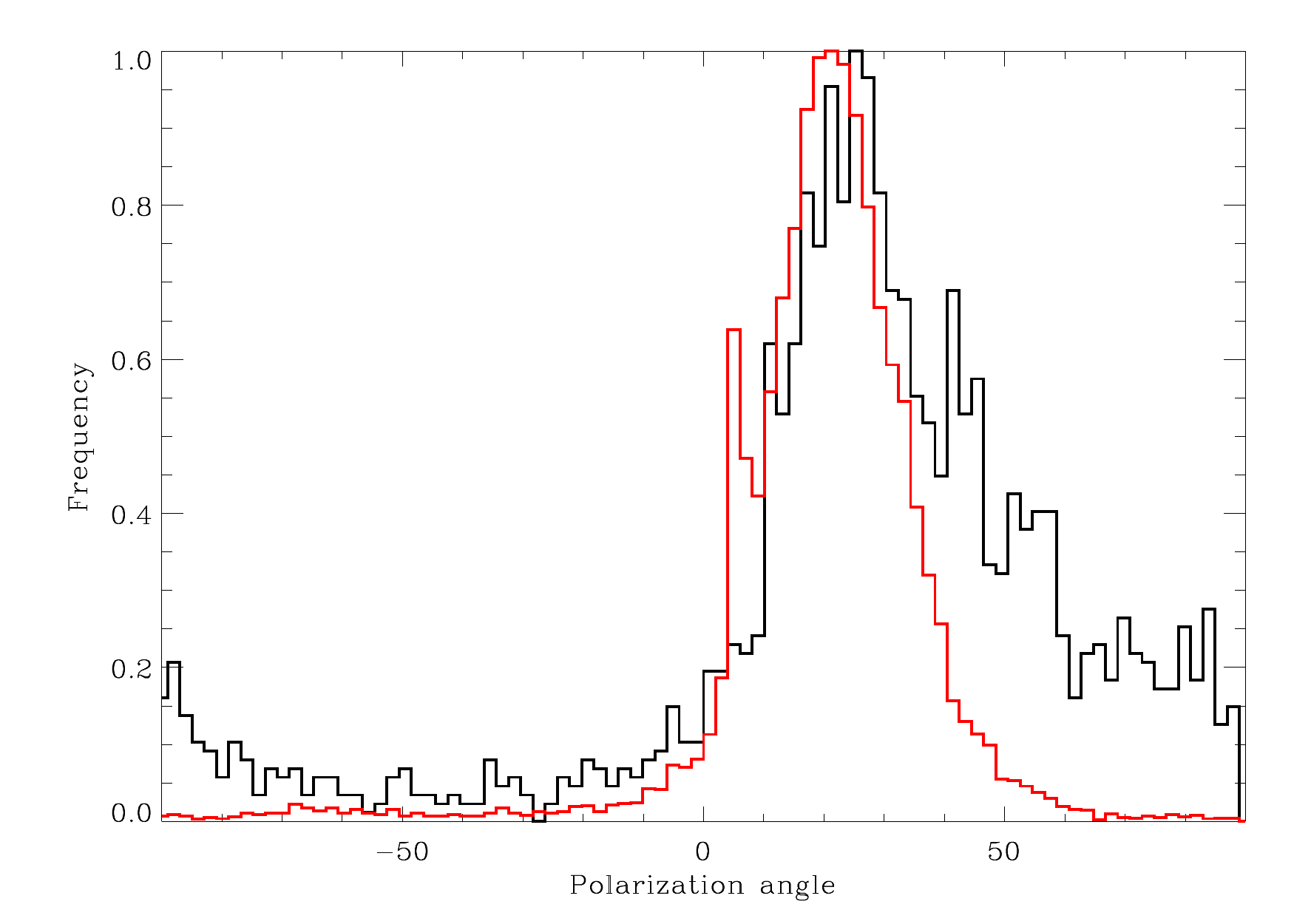}
	\caption{Normalized histograms of the {\PILOT} polarization angles obtained by analyzing the polarization angles map generated with the {\ROMA} map making (black) and with the {\Scanamorphos} map making (red). }
	\label{fig:polarization_angle_hist_roma_scanam}
\end{figure}

Fig. \ref{fig:polarization_angle_hist_roma_scanam} shows the $\psi$ map pixel histograms obtained with the {\scanamorphos} software and the {\ROMA} software in the signal dominated region defined by $-1.4\,^\circ<l<1.8\,^\circ$ and $-0.1\,^\circ<b<0.4\,^\circ$. 
, $\psi_0^\textsc{Roma}=18.37\,^\circ$)
, however the {\ROMA} histogram shows a larger dispersion which is due to the poorer quality of the {\ROMA} polarization maps that show residual stripes, in particular in the $Q$ map. As mentioned in Sec.~\ref{Sec:Roma}, this is due to the fact that the {\ROMA} map-making strongly depends on the accurate modeling of the detector noise and the noise covariance matrix for the moment does not include the off-diagonal terms linked to the correlated noise between the detectors. 

The very good agreement of the polarization angles obtained with two pipelines that rely on very different assumptions and are implemented independently shows the robustness of the results and it is an important test of the quality of the data.

\subsection{Comparison with {\Planck}}
\label{sec:comparison_planck}

The {\Planck} satellite experiment observed the whole sky with two instruments, the Low Frequency Instrument (LFI) from  30 to 70\,GHz and the High Frequency Instrument (HFI) from 100 to 857\,GHz. Among the nine observing bands, seven channels between 30 and 353\,GHz measured polarization in addition to intensity with spatial resolutions ranging from 30 to 5\,arcmin \citep{Planck_overview_2018}. In this paper, we make use of the publicly available {\Planck} 2018 release LFI and HFI maps, described in \citet{Planck_LFI_2018} and \citet{Planck_HFI_2018}. {\Planck} polarization maps adopt the {\sc HEALPix} projection scheme \citep{healpix}, in Galactic coordinates, and use the COSMO convention for the definition of the polarization angles. Polarization angles for Planck data are computed from the $Q$ and $U$ maps according to equation~\ref{eq:thetam}.

Given that dust emission is optically thin at {\Planck} frequencies as well as 240\,\micron, even towards the inner regions of the Galactic plane and that the same population of dust grains is expected to dominate the emission in both wavelength regimes, it is expected that the orientation of the polarization detected with {\Planck} and {\PILOT} should be similar. Different emerging polarization angles could result from emission by specific dust populations with different alignment properties or spatial distribution and varying $B$-field orientation along the line-of-sight. More probably however, differences would result from remaining systematic effects in the {\PILOT} or {\Planck} data. Comparing apparent polarization directions is therefore an important test of the data quality.

\begin{figure*}[ht]
\includegraphics[width=1.0\textwidth]{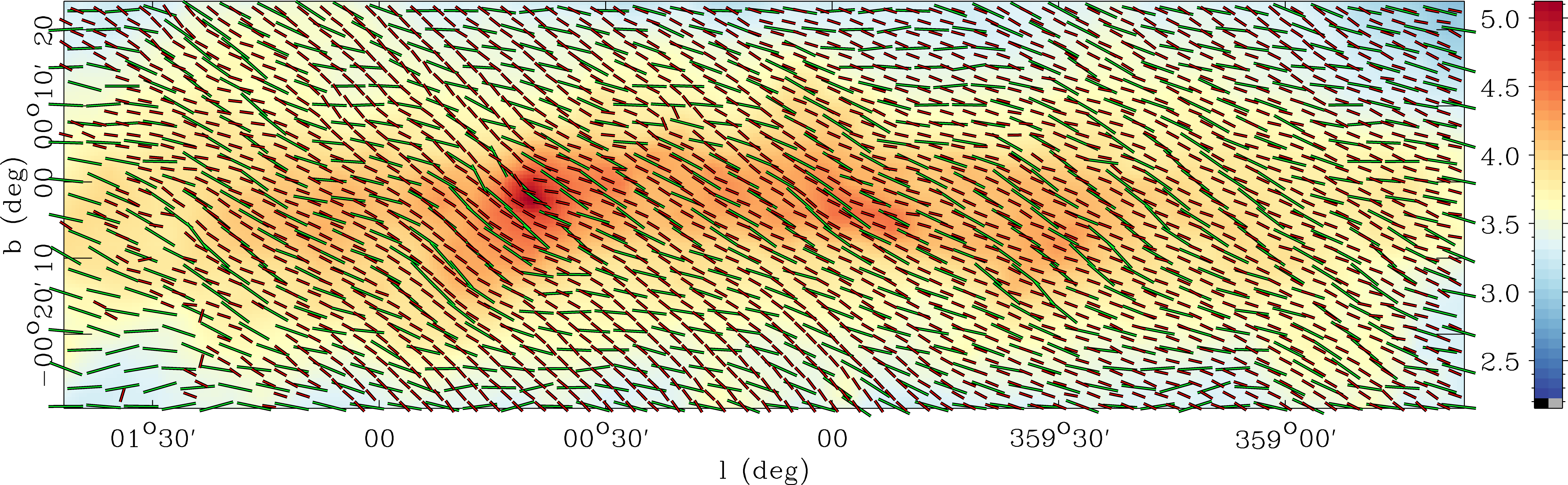}
\includegraphics[width=1.0\textwidth]{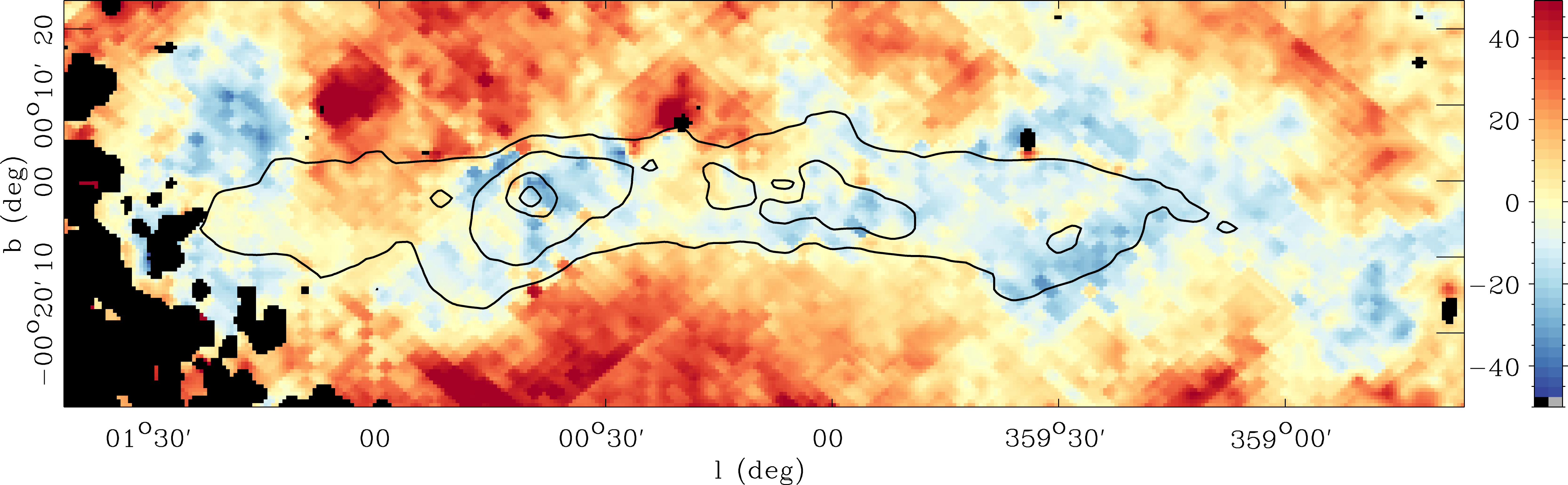}
\caption{\label{fig:polar_vector_Planck353vsPilot} Upper panel: Comparison between the $B$-field orientation as derived from the {\PILOT} 240\,\micron\ (red) and {\Planck} 353\,GHz (green) dust polarization measurements. The $B$-field orientation is shown as constant length line segments. The background image is the {\PILOT} Stokes $\StokesI$ map shown in log scale. Lower panel: map of the angle difference (in degrees) between {\Planck} 353\,GHz and {\PILOT} 240\,\micron.}
\end{figure*}

Figure~\ref{fig:polar_vector_Planck353vsPilot} compares the $B$-field orientation measured with {\PILOT} at 240\,\micron\ and {\Planck} at 353\,GHz. We can see, despite the different resolutions (5\,arcmin for {\Planck} 353\,GHz and 2 arcmin for {\PILOT}), that there is an overall agreement between the measured polarization directions. It can be noticed on this figure that the {\Planck} polarization angle directions are less homogeneous than the directions measured by {\PILOT}. The difference map of the polarization angle between {\Planck} and {\PILOT} (bottom panel) shows some excursions as large as 45 degrees for some pixels, but are spread around zero, especially along the Galactic plane. We can see on this difference map obvious imprint of the low $N_\mathrm{side}$ {\sc HEALPix} pixels that can be seen in the {\Planck} 353\,GHz polarization angle map (and can not come from {\PILOT} maps as they are not projected using {\sc HEALPix}).  

\begin{figure}[!ht]
	\centering
	\includegraphics[width=0.5\textwidth]{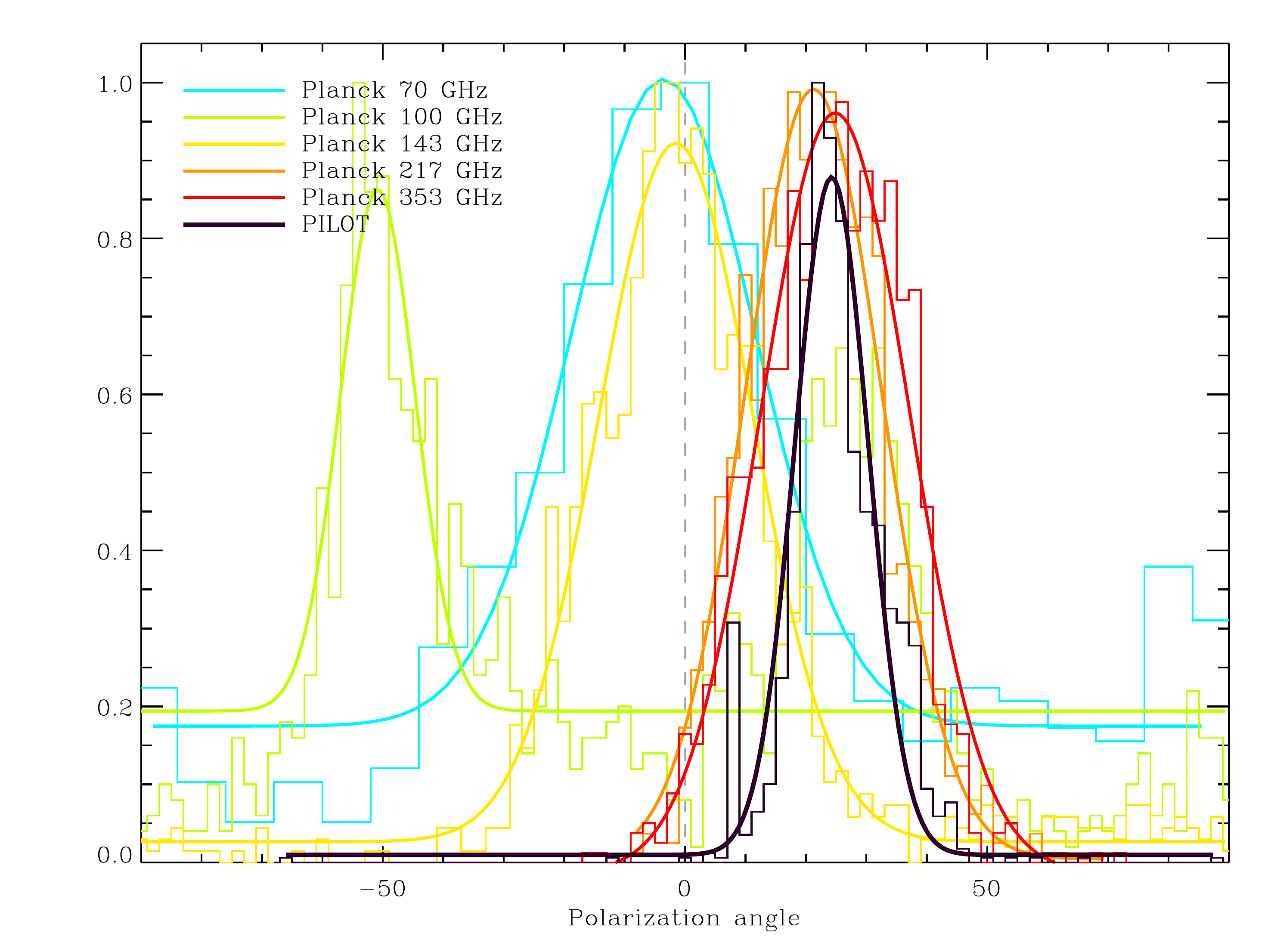}
	\caption{Histogram of the {\PILOT} polarization angles (black) on the galactic center region compared to {\Planck} observations at 70, 100, 143, 217 and 353\,GHz (blue, green, yellow, orange and red, respectively). We overplot the Gaussian fits of these histograms as thicker lines to guide the eye.}
	\label{fig:polarization_angle_hist}
\end{figure}

Figure~\ref{fig:polarization_angle_hist} shows the histograms of the polarization orientation observed with {\PILOT} (degraded to a resolution of 5 arcmin, to match {\Planck} highest frequencies') and {\Planck} (70, 100, 143, 217 and 353\,GHz). Histograms are computed from the polarization maps inside a box having $-1.3\,^\circ<l<1.8\,^\circ$ and $-0.4\,^\circ<b<0.1\,^\circ$, corresponding roughly to where the intensity measured by {\PILOT} at 240\,\micron\ is larger than 500\,MJy/sr. The histogram of the {\PILOT} polarization angle peaks at a value (the Gaussian best fit of the histogram has $\psi_0=24.1\,^\circ$, $\sigma=5.9\,^\circ$) very similar to where {\Planck} 217 and 353\,GHz histograms peak ($\psi_0=21.3\,^\circ$, $\sigma=11.3\,^\circ$ and $\psi_0=24.9\,^\circ$, $\sigma=12.3\,^\circ$, respectively), showing that {\PILOT} sees the same dust as {\Planck} at 217 and 353\,GHz. The {\PILOT} histogram is narrower than the {\Planck} ones (when {\PILOT} data resolution is degraded to 5\,arcmin), highlighting the more homogeneous polarization directions already observed in Fig.~\ref{fig:polar_vector_Planck353vsPilot}. 

The {\Planck} 100\,GHz channel is expected to have an important CO-line contribution towards the Galactic center region and therefore a strong spurious polarization due to spectral mismatch between bolometers. This spurious polarization was corrected for in the {\Planck} 2018 maps, but the correction is known to be inaccurate in this region were the matter velocity can shift the CO-line significantly with respect to other regions of the sky, where the 100\,GHz bolometer response to the CO-line was calibrated for spurious polarization correction \citep{Planck_HFI_2018}. We interpret the mean value of the angle at 100\,GHz of $\sim-50\,^\circ$ to be partially due to this residual spurious polarization. At the other Planck frequencies (70 and 143\,GHz), the histograms peak around zero degrees ($\psi_0=-3.6\,^\circ$, $\sigma=15.2\,^\circ$ and $\psi_0=1.5\,^\circ$, $\sigma=12.8\,^\circ$, respectively).

\begin{figure*}[!ht]
	\centering
	\includegraphics[width=1.0\textwidth]{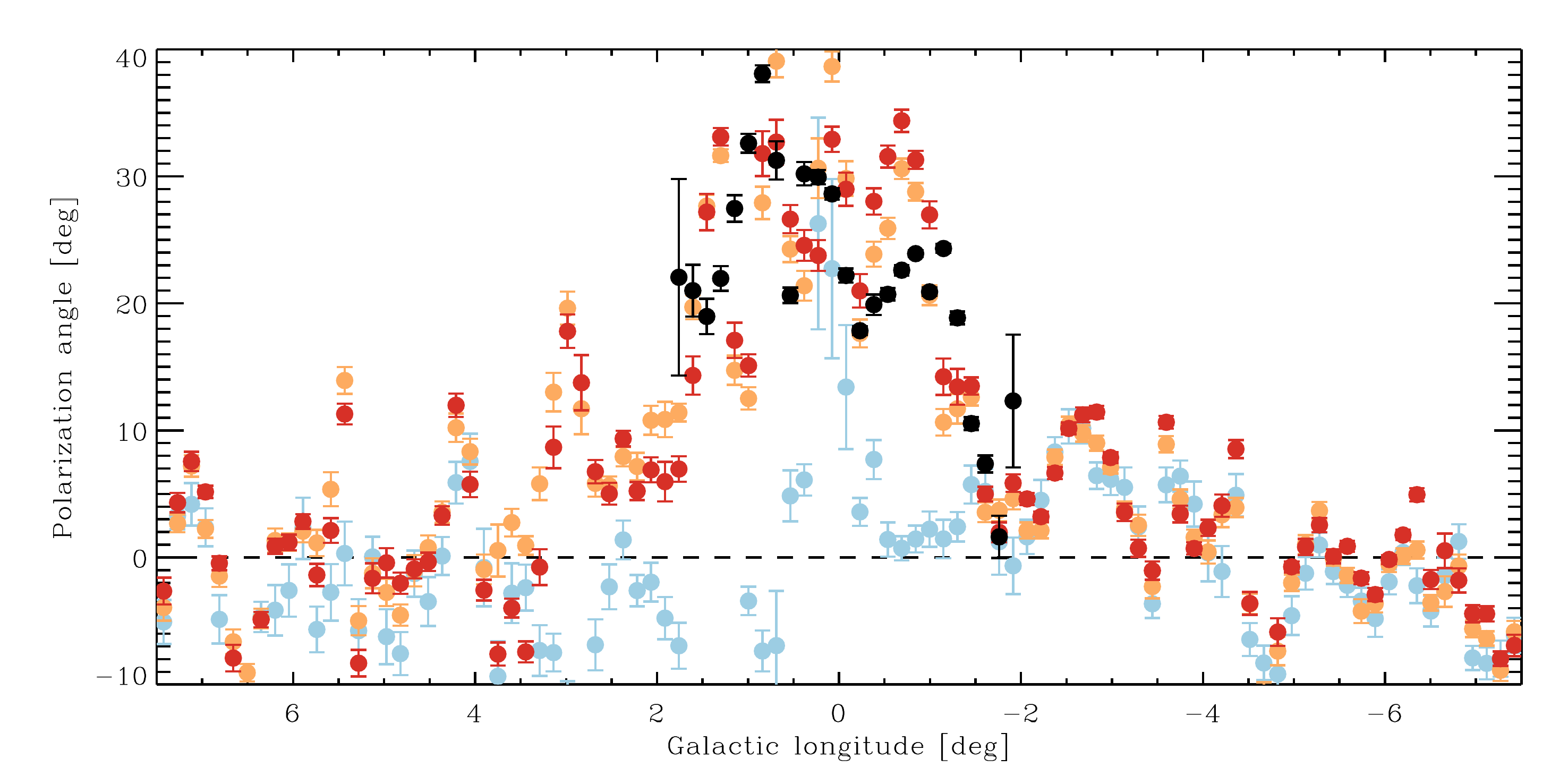}
	\caption{Polarization angle profile as a function of Galactic longitude. The mean polarization angle is computed from longitudes between -7.5$\,^\circ$ and 7.5$\,^\circ$ and latitudes between -0.15$\,^\circ$ and 0.15$\,^\circ$ in 100 boxes of size 0.15$\,^\circ$ in longitude. The longitude profiles are shown for the {\Planck} 143, 217 and 353\,GHz channels (blue, orange and red, respectively) and {\PILOT} (black). The error bars represent the standard deviation of the angle in each boxes.}
	\label{fig:polarization_longitude}
\end{figure*}

In Fig.~\ref{fig:polarization_longitude}, we look at the Galactic longitude profile of the {\PILOT} and {\Planck} polarization angles, computed in boxes of size $\Delta l=0.15\,^\circ$ and $\Delta b=0.3\,^\circ$ on either side of the Galactic plane, ranging from $l=-7.5\,^\circ$ to $l=7.5\,^\circ$. We can see that the {\PILOT} data polarization angle profile follows fairly that of {\Planck} 217 and 353\,GHz and that the discrepancies between {\PILOT} and {\Planck} 217 and 353\,GHz is of the same order than that of the discrepancies between those two {\Planck} channels. The global dust polarization profile, traced by {\PILOT} and {\Planck} 217 and 353\,GHz, shows a polarization angle close to zero (magnetic field projection on the sky along the Galacitc plane) from $l=-7.5\,^\circ$ to $l=7.5\,^\circ$, except for a region corresponding to the extent of the CMZ (from $-1\,^\circ$ to $1.5\,^\circ$ \citep{Molinari_etal2011}) where the measured angles range from 20 to $30\,\degr$. At 143\,GHz, the polarization angle follows that of the dust outside the CMZ ($l<-1\,^\circ$ and $l>1.5\,^\circ$) but shows completely different behaviour inside, being much closer to zero (as noticed in Fig.~\ref{fig:polarization_angle_hist}). This is the hint for another component than dust -- presumably synchrotron -- dominating the emission in the CMZ at frequencies under 143\,GHz. 

\section{Discussion}\label{sec:discussion}

\subsection{Do {\PILOT} polarization measurements probe the field structure in the CMZ?}
\label{sec:cmz}

Analysis of the {\Planck} satellite dust polarization data has shown that depolarization due to the complex nature of the field structure is omnipresent (see e.g. \cite{Planck2015_intermediate_XIX}). This is especially true towards the Galactic plane where the magnetic field is likely to rotate significantly along the line of sight and within the instrument beam as result of variations of the ordered field and turbulence. Even if dust emission is optically thin at the observed frequencies, it is thus not necessarily the case that polarization measurements are sensitive to the emission along the whole line-of-sight, as the signal may suffer significant depolarization.

\noindent Here we verify that when observing towards the Galactic center with {\PILOT}, the polarized signal is indeed dominated by the CMZ region. For this, we use a simplified model of the ordered and turbulent distribution of the magnetic field in the Galactic disk. For the dust density distribution, we assume that the Milky Way has a four-armed spiral structure, superimposed on an axisymmetric exponential profile. The model, which is described in Appendix~\ref{sec:gal_model_annex}, is constrained using the observed polarization fraction profile along the Galactic plane obtained from the {\Planck} RC4 data. We use this model to estimate the relative contribution of structures in the Galactic disk along the line of sight towards the region observed with {\PILOT}. We explore three representative regimes: no turbulent field, equipartition between the ordered and turbulent fields, and a turbulent field that is three times stronger than the ordered field. In all cases, turbulence in the disk along the line-of-sight towards the Galactic Center is insufficient to depolarize the dust emission to the low values ($\polfrac \sim 1$ to 2\,\%) that are observed.  We conclude that the observed polarization properties are due to emission that is intrinsic to the CMZ. This is qualitatively confirmed by the good correspondence between emission structures in the {\PILOT} total intensity map and the $\StokesQ$, $\StokesU$ and polarized intensity ($P$) maps of the CMZ (see Fig.~\ref{fig:IQU_scanam_zoomed}).  

\subsection{Uniform orientation of the POS {\rm B}-field}
\label{sec:homogeneity}

\noindent The orientation of the plane of the sky (POS) magnetic field inferred from the polarization direction measured by {\PILOT} is surprisingly uniform across the $3.1\,\degr \times 0.9\,\degr$ field presented in this paper (see Sect.~\ref{Sec:Polarization_results}). In this subsection, we quantify the regularity of the field on linear (and not just angular) scales in the CMZ via a comparison with the dispersion in the dust polarization angles in the local ISM measured by Planck. We then consider how the observed regularity of the field can be reconciled with the low fraction of dust polarization that is observed towards this region by Planck, and what we can infer about the large-scale geometry of the field in the Galactic Center region.

\noindent The polarization angle dispersion function $\DeltaAng$ quantifies how polarization angles decorrelate with increasing angular scale (see Eq.~\ref{equ:delta_phi1}). In \cite{Planck2015_intermediate_XIX}, the large scale distribution of $\DeltaAng$ was studied at an angular resolution of $1\,\degr$ and associated lag of $30$\,arcmin, since these parameter values mitigate noise bias for an intermediate latitude study.
Using the same {\Planck} data product, working resolution and lag as in \cite{Planck2015_intermediate_XIX}, we find a median value of $\DeltaAng^{353}(\lag=30\,\arcmin)=3.2\,\degr$ in the region covered in Fig.\,\ref{fig:IQU_scanam_zoomed}. This value is close to the median value of $\DeltaAng$ measured over the entire mid-latitude sky by \cite{Planck2015_intermediate_XIX}. We emphasize, however, that this apparent agreement does not mean that the field in the Galactic Center and local ISM have similar structural properties. The dust emission measured by {\Planck} at intermediate latitudes is produced in the relatively nearby ISM (within $\sim 1\,\mathrm{kpc}$). An angular scale measured at intermediate latitude  therefore corresponds to a spatial scale that is $\sim8$ times smaller than the spatial scale corresponding to the same angular scale measured at the Galactic center. Expressed otherwise, our measurement of $\DeltaAng^{353}(\lag=30\,\arcmin)=3.2\,\degr$ in the Galactic Center indicates that the POS field structure remains homogeneous over a $\sim8$ times larger length scale than in the local ISM.

\noindent
We constructed a map of $\DeltaAng$ from the {\PILOT} data itself. After degrading the resolution of the {\PILOT} data to $5\,\arcmin$ to mitigate noise bias, we obtain a median value of $\DeltaAng^{240}(\lag=2.5\,\arcmin)=2.5\,\degr$.
This value is consistent with the median $\DeltaAng(\lag=30\,\arcmin)$ from {\Planck}, considering that it is measured at a smaller lag. We conclude that the POS magnetic field orientation in the GC region is intrinsically more regular than in the nearby ISM.

\noindent The analysis of the {\Planck} data demonstrated that there is an anti-correlation between $\DeltaAng$ and dust polarization fraction $\polfrac$ \citep[see Fig. 23 of][]{Planck2015_intermediate_XIX}, a relation that can be used to infer the characteristic value of $\polfrac$ that would be expected for a given value of $\DeltaAng$. For $\DeltaAng^{353}(\lag=30\,\arcmin)=3.2\,\degr$, the $\DeltaAng$-$\polfrac$ anti-correlation would predict $\polfrac=6.2\,\%$. The largest possible $\polfrac$ values compatible with this $\DeltaAng$ value -- corresponding to an orientation of the ordered $B$-field that is maximally in the plane of the sky -- would be very close to $\polfrac\simeq20\,\%$, the maximum value observed on the sky in the {\Planck} data at 353\,GHz \citep{Planck2015_intermediate_XIX, Planck_polar_2018}. However, the polarization fraction at 353\,GHz that is actually measured by {\Planck} in the Galactic Center region is $\polfrac=1.6\,\%$ only.
This leads to a value for the product $\DeltaAng \times \polfrac=0.05\,\degr$, much smaller than observed values in the {\Planck} data, including towards regions with column densities similar to those of the CMZ \citep[see in][]{Planck_polar_2018}.

\noindent Possible explanations for the low polarization fraction in the Galactic Center region include (1) the Galactic Center magnetic field has a strong fluctuating component, (2) dust grains in the Galactic Center are more weakly aligned with the magnetic field than elsewhere in the Galaxy and (3) the Galactic Center magnetic field has a significant LOS component. The first possibility seems to be ruled out by the observed regularity of the POS magnetic field orientation. The second possibility cannot be discarded, but there is currently no observational evidence to support it. Here we explore the third scenario, i.e. that the observed homogeneity of the POS $B$-field orientation and the low dust polarization fraction can be self-consistently explained by a 3D ordered field that is oriented close to the LOS. Since $\polfrac \propto \sin^2 \gamma$
where $\gamma$ is the angle of the field to the LOS, the values of the dust polarization fraction discussed above ($6.23\,\%<\polfrac<20\,\%$) imply $15\,\degr<\gamma<30\,\degr$. If the distribution around the {\Planck} $\DeltaAng$-$\polfrac$ anti-correlation at constant $\DeltaAng$ is mostly due to variations in $\gamma$ along any LOS, then we favour $\gamma\sim15\,\degr$ for the Galactic Center, corresponding to a dust polarization fraction close to the intrinsic maximum value.

A mean magnetic field that is weakly inclined to the LOS in the GC region could possibly be explained by the geometry of the large-scale Galactic magnetic field. For instance, if the large-scale field is bi-symmetric and follows the Galactic bar, which is now believed to be inclined by $\sim 30\,^\circ$ with respect to the LOS \citep{blandhawthorn&g_16}.

The POS magnetic field in the GC region is found to be tilted clockwise to the trace of the Galactic plane by $\simeq +22\,^\circ$ on average.
Fig.\,\ref{fig:polarization_longitude} shows the polarization angle profile as a function of Galactic longitude for  the {\Planck} 143, 217 and 353 GHz channels (blue, orange and red, respectively) and {\PILOT} (black). The mean polarization angle is computed from longitudes between -7.5$\,^\circ$ and 7.5$\,^\circ$ and latitudes between -0.15$\,^\circ$ and 0.15$\,^\circ$ in 100 boxes of size 0.15$\,^\circ$ in longitude. The error bars represent the standard deviation of the angle in each boxes. The 
comparison with larger-scale {\Planck} maps indicates that the $+22\,^\circ$ tilt angle is actually restricted to the longitude range $-1.5\,\degr<l<+1.5\,\degr$. Outside this range, the tilt angle approaches zero, i.e., the POS magnetic field becomes approximately parallel to the Galactic plane. 

The measured tilt angle $\simeq +22\,^\circ$ is not easily understood. There is no obvious gaseous structure that fills the GC region and has a $+22\,^\circ$ clockwise tilt to the Galactic plane. The CMZ spans roughly the above longitude range, but it is much less extended in latitude, and only a relatively small portion of it (the western part of the twisted ring; see Sect.~\ref{Sec:infinityloop}) was observed to have a clockwise tilt compatible with $+22\,^\circ$.

\subsection{Characteristic magnetic field strength in the CMZ}\label{sec:B_strength}

The regularity of the POS magnetic field orientation in the {\Lzero} region suggests that the magnetic field there is very strong, i.e., strong enough for magnetic pressure, $P_\mathrm{mag}$, to withstand the ram pressure from  ambient interstellar clouds, $P_\mathrm{ram}$. Using a conservative estimate for $P_\mathrm{ram}$ in the Galactic Center region, \cite{yusef&m_87} previously showed that the condition $P_\mathrm{mag} \gtrsim P_\mathrm{ram}$ is roughly equivalent to $B \gtrsim 1\,\mathrm{mG}$.

\noindent Here, we employ similar reasoning in an effort to estimate the magnetic field strength using the observed structure of the POS magnetic field in the CMZ region. As described in Section~\ref{sec:pilot_maps}, we observe that the POS magnetic field orientation tends to become more parallel to the Galactic Plane in regions of high brightness. This is especially evident for the high-brightness structure in our total intensity map corresponding to Sgr B2, the 20\,\kms\ and 50\,\kms\ clouds and the Brick, which exhibit significant departures from the mean POS magnetic field orientation $\simeq +22\,^\circ$. We can thus reasonably imagine that these clouds have a ram pressure perpendicular to the magnetic field that is comparable to the local magnetic pressure: $P_\mathrm{ram,\perp} \sim P_\mathrm{mag}$, where $P_\mathrm{ram,\perp} = \rho_\mathrm{cl} \ \delta v_\perp^2$, $\rho_\mathrm{cl}$ is the mean mass density of the cloud, and $\delta v_\perp$ is the relative cloud velocity (with respect to its surroundings) perpendicular to the magnetic field. 

\noindent For the characteristic gas densities in these regions $\rho_\mathrm{cl}$, we adopt values obtained from HC$_{3}$N excitation studies by \citet[][]{Mills_etal2018} and \citet[][]{Morris_etal1976}.  For the LOS velocity of a cloud $\delta v_\mathrm{LOS}$ relative to its surroundings, we use the kinematic structure of the CMZ region presented by  \citet{Henshaw_etal2016}. The  values that we adopt for each cloud are listed in Table~\ref{tbl:bfield_estimates}. To obtain the corresponding real space (i.e. 3D) relative velocities, we assume that the clouds as well as their surroundings follow the orbital trajectory proposed by \citet{Kruijssen_etal2015}. Finally, we project the 3D relative velocities onto the plane perpendicular to the mean magnetic field, which we take to be inclined by $15\,^\circ$ to the LOS and to have a POS orientation angle of $+22\,^\circ$ (see Sect.~\ref{sec:homogeneity}). With these parameter values, we obtain magnetic field strength estimates ranging between 1.4 and 4.7\,mG as shown in  Table~\ref{tbl:bfield_estimates}. 

Our four magnetic field strength estimates are reasonably close to each other. The spread by a factor $\sim 3$ can be attributed to differences in the estimated ram pressures of the four clouds. As expected, clouds with larger ram pressures tend to cause stronger perturbations in the magnetic field. Our field strength estimates are also very high by interstellar standards. Indeed, $B$ is typically $\sim$ a few $\mu\mathrm{G}$ in the Galactic disk and almost certainly stronger in the Galactic center region, with estimates ranging from $\sim 10\,\mu\mathrm{G}$ \citep{larosa_etal_05} to $B \gtrsim 1~\mathrm{mG}$ \citep{yusef&m_87} (see \citealt{ferriere_09} for a review).

\begin{table}[]
    \centering
    \begin{tabular}{lccc}
               \hline
         Cloud & $\log(n_{\mathrm{H}_2})$ & $\delta v_\mathrm{LOS}$ & B \\
               & $[\mathrm{cm}^{-3}]$ & $[\kms]$ & $[\mathrm{mG}]$ \\
               \hline
Sgr B2  & 5.0  &  20  &      4.7 \\
Brick  &  4.2  &  5  &      1.4 \\
50\kms & 4.5  &  10  &      2.7 \\
20\kms &   4.3  &  10  &      2.8 
    \end{tabular}
    \caption{Magnetic field strength estimates for CMZ clouds, assuming equipartition between magnetic pressure and ram pressure (see text in Section~\ref{sec:B_strength}).}
    \label{tbl:bfield_estimates}
\end{table}

\subsection{Coincidence with the 100-pc ring}
\label{Sec:infinityloop}

\begin{figure}
\includegraphics[width=0.45\textwidth]{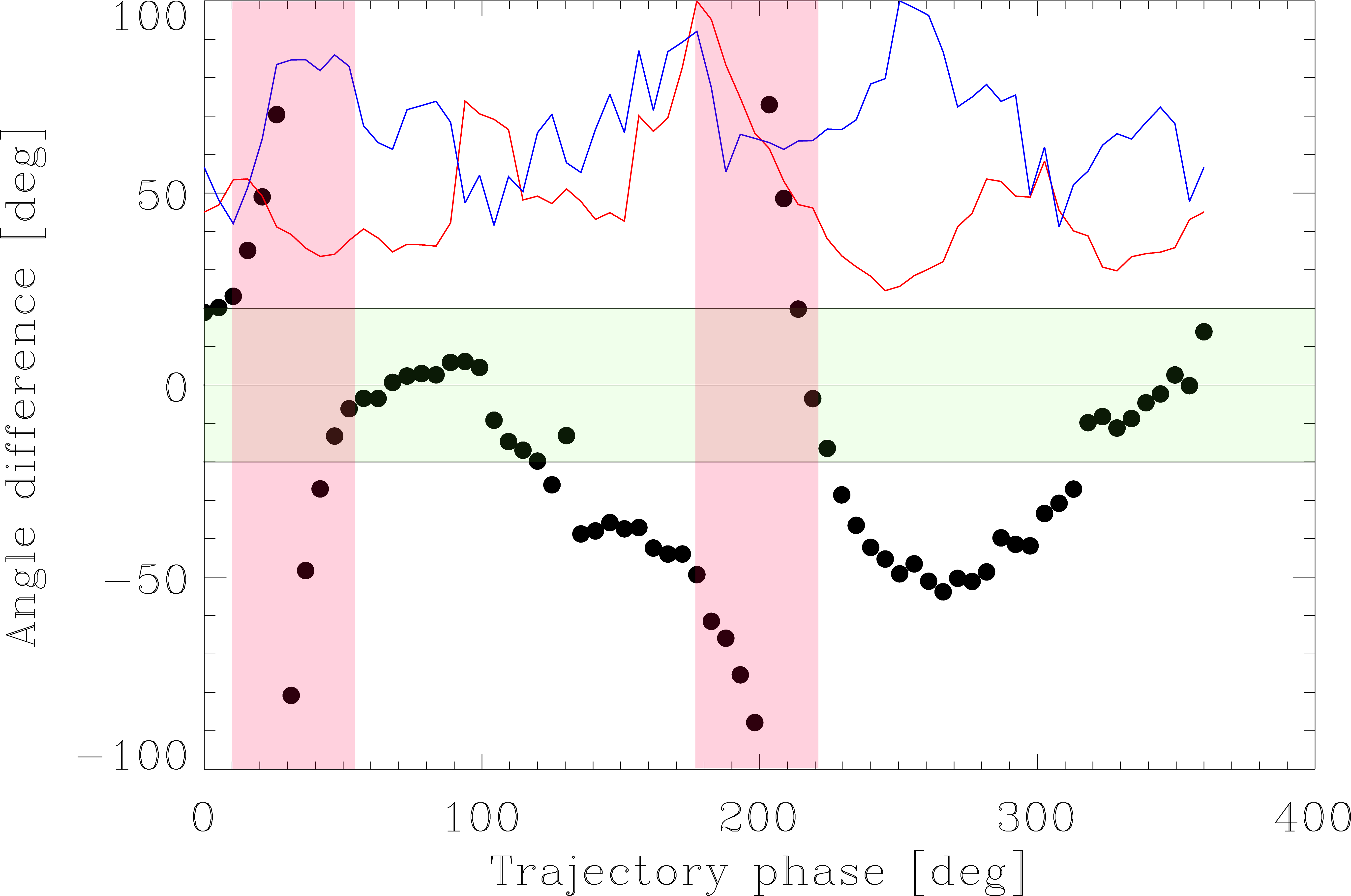}
\caption{Angle difference between the $B$-field orientation and the tangent to the 100\,pc ring as defined in \cite{Molinari_etal2011} as a function of the phase angle along the ring (black dots). The red and blue profiles show the corresponding $240\,\mic$ intensity and polarization fraction profiles in arbitrary units.
The vertical lines show the range $\pm 20\,\degr$.
The vertical boxes show the tangent point of the ring.
}
\label{fig:Molinari_profiles}
\end{figure}

The molecular cloud at the center of the {\CMZ} is likely to orbit around the {\GC} following well defined trajectories, as described in \cite{Molinari_etal2011} and \cite{Kruijssen_etal2015}. A legitimate question is to ask if the magnetic field as traced by dust associated to those clouds follows these structures even partly, or if the field lines are affected by the gas motion along those trajectories.
If the field lines are significantly dragged by the gas motion, we would expect the $B$-filed orientation to be somewhat parallel to the trajectories. We would also expect the polarization fraction to lower at the tangent point of the trajectories, where the $B$-field orientation would be essentially along the LOS.
We used the analytical description of the trajectory proposed in  \cite{Molinari_etal2011} and the tabulated orbital values in \cite{Kruijssen_etal2015} to construct the sky projected trajectory track on the plane of the sky. This track is overlaid for the \cite{Molinari_etal2011} 100-pc ring on Fig.~\ref{fig:schematic}. We then compute the tangent direction to the track at each location along the trajectory. We finally compute the angle difference between this direction and the $B$-field orientation as measured by our polarization data, at each location along the track.

Figure\,\ref{fig:Molinari_profiles} shows the angle difference between the $B$-field orientation as measured with {\PILOT} and the tangent to the 100-pc ring. There is a very decent match between the $B$-field direction and the tangent to the trajectories of the Molinary 100-pc ring over half of the trajectory spanning negative longitudes. Over this section of the track, the $B$-field follows the tangent to the track to better than $8\,\degr$. Those locations are indicated in Fig.\,\ref{fig:schematic}. Along the location at positive longitudes however, the match is poor, with angle differences approaching $\simeq 45\,\degr$.
Using the physically motivated trajectories from \cite{Kruijssen_etal2015} does not lead to a better overall match.
In both cases, we looked but did not find any clear evidence for a trend for a decrease polarization fraction at the tangent points of the trajectories.
These analyses confirm the visual impression when comparing the field orientation of Fig.\,\ref{fig:polar_vector_scanam} and the trajectories in Fig.\,\ref{fig:schematic} that the magnetic field orientation is very homogeneous over the whole region, as described in Sect.\,\ref{sec:homogeneity}, and shows very little direction change at the location of the trajectories. Therefore, the match in direction over half the 100-pc ring as described in \cite{Molinari_etal2011} could also be purely fortuitous.

\subsection{Comparison with previous ground measurements}\label{sec:ground_measurements}

\begin{figure}
\includegraphics[width=0.45\textwidth]{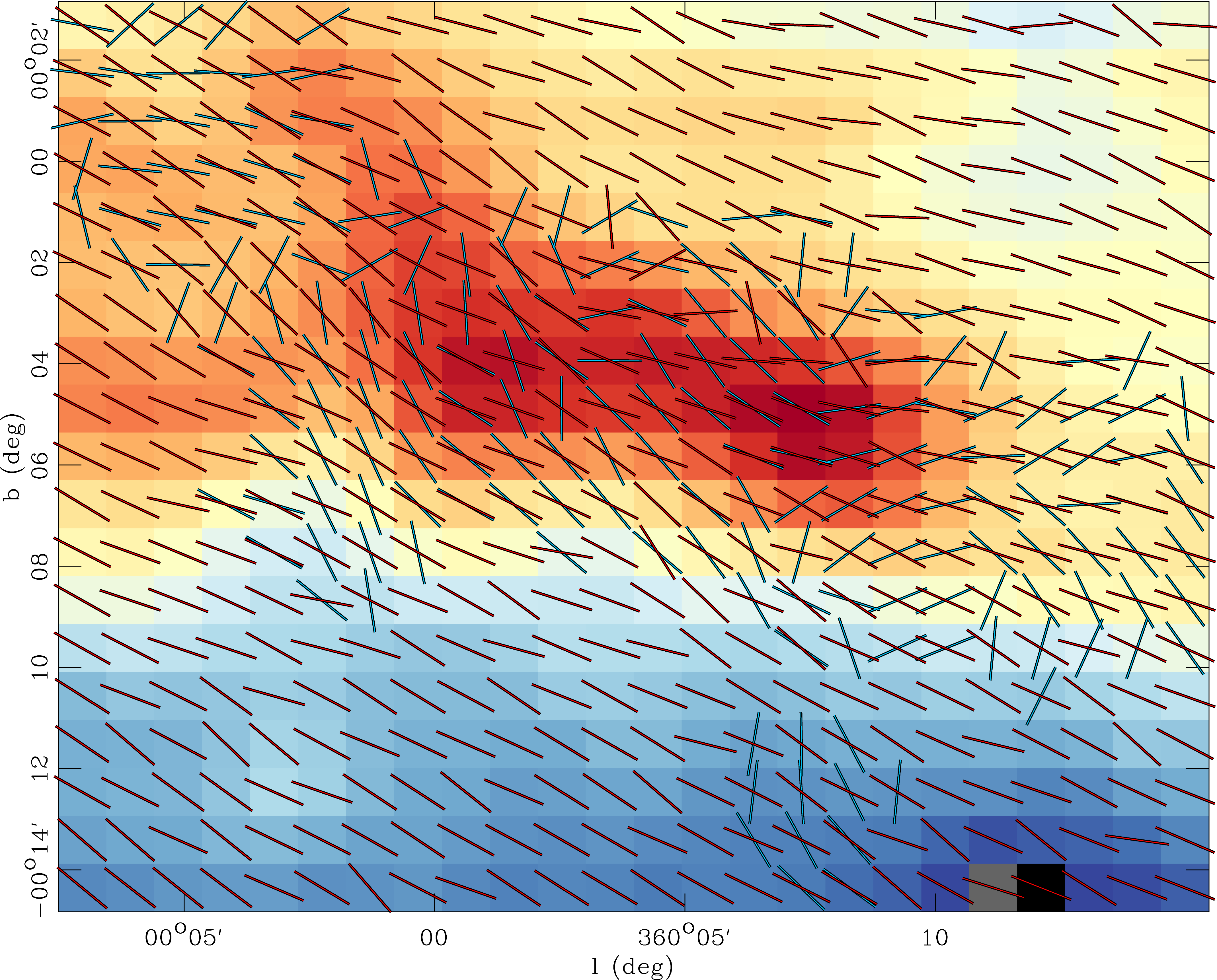}
\includegraphics[width=0.45\textwidth]{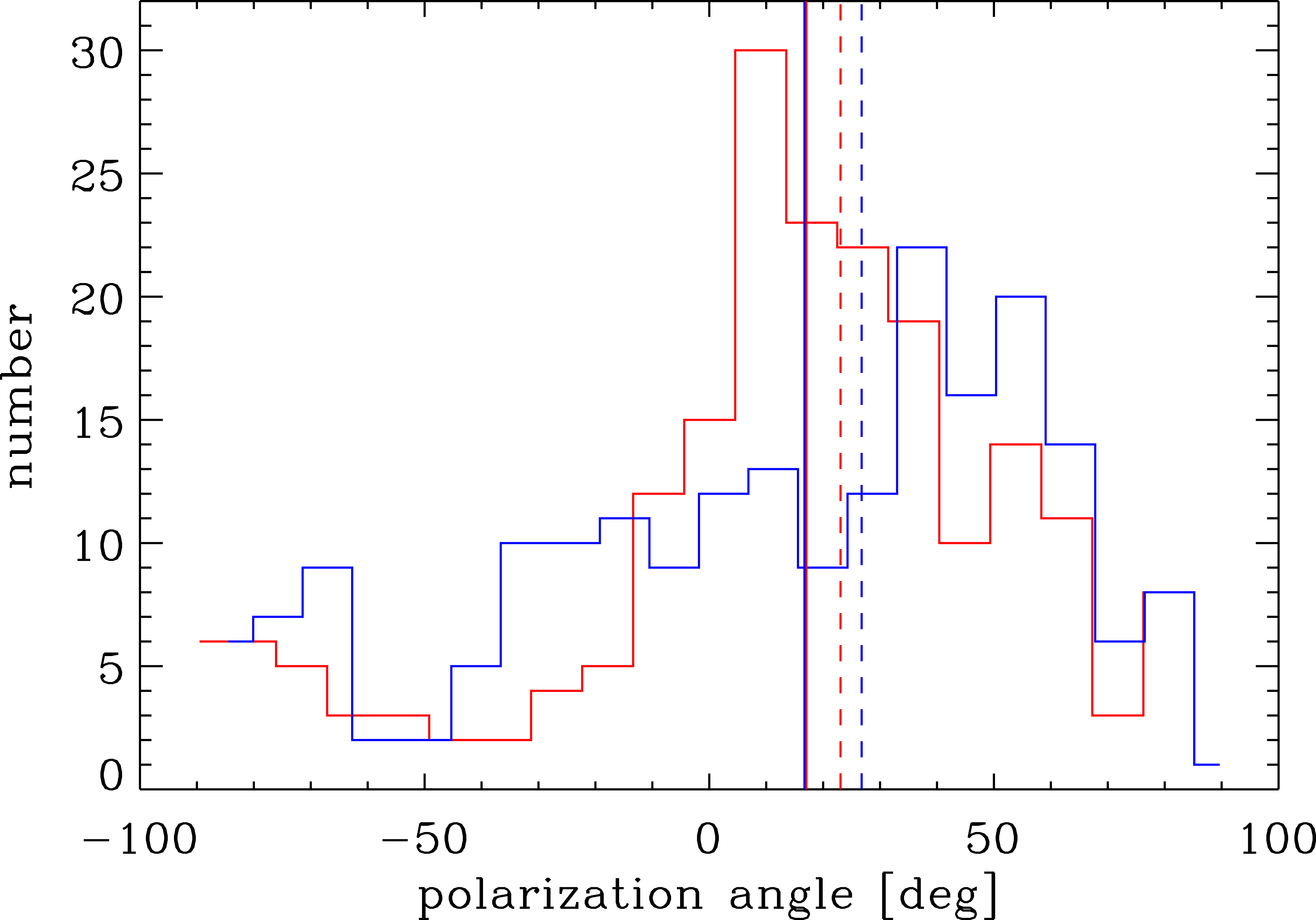}
\caption{Upper panel: $B$-field vector of the averaged JCMT data (blue) and {\PILOT} background subtracted data (red) over the extent of cloud 50\kms covered by the JCMT data, overlaid on the {\PILOT} intensity image.
Lower panel: Histogram of the JCMT (blue) and {\PILOT} (red) $B$-field orientation. The vertical lines show the mean (solid) and median (dashed) angle values of each distributions.
}
\label{fig:pilot_vs_jcmt}
\end{figure}

In principle, the $B$-field orientation inferred from our data could be compared to that derived from similar FIR/submm data obtained from ground observations, as long as these observations are dominated by polarized emission from large dust grains. In practice, however, the comparison is complicated by several limitations. First, ground data are often obtained at high angular resolution, but cover very small sky regions. As a consequence, the data tabulated in the literature is often too scarce to fill-up one the {\PILOT} beam, preventing the comparison. Second, the data processing of ground data is likely to filter large scale emission. A fair comparison would require filtering the {\PILOT} data using the transfer function of the data processing applied to the ground data, but this is usually not feasible, since the information about the processing is often incomplete and the processing software is not readily available.

Out of the FIR/submm polarization data available described in Sect.~\ref{sec:Intro}, we selected the JCMT data of \cite{Matthews2009} obtained at $450\,\mic$ as the one with spatial coverage and completeness best matching our resolution. This data covers most of the extent of the 50\kms molecular cloud highlighted in Fig.~\ref{fig:schematic}.
Figure\,\ref{fig:pilot_vs_jcmt} shows the comparison between the JCMT data and the {\PILOT} data. In order to perform the comparison, we transformed the JCMT data provided under the form of polarization fraction and angle into Stokes parameters and we computed the total intensity weighted average of the JCMT  Stokes parameters falling in each pixel of our map. Those values were sky rotated to the same angle convention and coordinate system as the {\PILOT} data. We then computed the corresponding angle orientation $\polang_\textrm{JCMT}$. In a simplified attempt to reproduce the low spatial frequency filtering of the signal in the JCMT data, we subtracted from the {\PILOT} $\StokesQ$ and $\StokesU$ maps the average values measured in the region of Fig.\,\ref{fig:pilot_vs_jcmt} covered by our data but not observed in the JCMT data. The background subtracted map was used to compute $\polang$. 
The top panel of Fig.\,\ref{fig:pilot_vs_jcmt} shows the $B$-field orientation of JCMT and {\PILOT}, overlaid to the {\PILOT} intensity map, within the cloud 50\kms region  covered by the JCMT data.
The bottom panel of Fig.\,\ref{fig:pilot_vs_jcmt} shows the histogram of the angles obtained from the two datasets. Although there is a significant dispersion and the statistics are low, the average of the angles are $17.05\,\degr$ and $16.72\,\degr$ for the {\PILOT} and JCMT data respectively, and the corresponding average angle difference between the two datasets is $0.54\,\degr$. Note that, if no background subtraction is applied to the {\PILOT} data, the histogram is significantly skewed towards a non-zero angle difference value. This indicates that the somewhat large dispersion is unlikely to be due to noise in the {\PILOT} or JCMT data, but is more probably due to inadequate filtering of the {\PILOT} data. A more accurate comparison would require processing the {\PILOT} data through the JCMT pipeline used by \cite{Matthews2009}, which is beyond the scope of this study.

The reasonable agreement between the observed $B$-field angles between the two experiments at different FIR frequency confirms that the main orientation of the $B$-field observed on large scales with {\PILOT} persist with a very similar orientation within the 50\kms molecular cloud.

\section{Summary and conclusions}
\label{sec:conclusions}

\noindent In this paper, we have presented new measurements of the polarized dust emission at 240\,\micron\ in the Galactic center region. The data were obtained during $\simeq 30$ minutes of observations during the second flight of the {\PILOT} balloon-borne experiment, which was launched from Alice Springs, Australia in April 2017. The observed region covers a wide field ($-2\,\degr<\glon<2\,\degr$, $-4\,\degr<\glat<3\,\degr$) at an angular resolution of $2.2\,\arcmin$. 
\noindent This paper is the first {\PILOT} paper to present scientific analysis of polarization measurements towards astronomical targets. To validate these measurements and assess the overall quality of our data processing pipeline, we present polarization maps from the {\PILOT} data using two independent map-making methods ({\scanamorphos} and {\ROMA}). The two pipelines generate similar $\StokesI$, $\StokesQ$ and $\StokesU$ maps, and yield a similar distribution of polarization angles across our observed field. The {\scanamorphos} maps of $\StokesI$, $\StokesQ$ and $\StokesU$ used for the analysis in this paper are publicly available  through the Centre d'Analyse des Donn\'ees \'Etendues (CADE, {\tt http://http://cade.irap.omp.eu}). 

\noindent The key results of our analysis are:
\begin{itemize}
    \item {We obtain a clear detection of polarized thermal dust emission from the Milky Way's Central Molecular Zone (CMZ). This is obvious from the observed structure of  emission features in the {\PILOT} $\StokesI$, $\StokesQ$ and $\StokesU$ maps (Figure~\ref{fig:IQU_scanam_zoomed}). We develop a Galactic model with ordered and turbulent magnetic field components and a realistic density distribution to demonstrate quantitatively that the observed polarization properties cannot be explained by the magnetic field of the Galactic disk along the line-of-sight to the Galactic Center region.}   
    \item {The {\PILOT} polarization angle measurements show that the structure of the magnetic field in the plane of the sky (POS) is remarkably uniform over the full extent of our mapped region. The observed distribution of polarization angles indicates that the POS $B$-field in the CMZ has a dominant orientation of
    $\simeq 22\,^\circ$ measured clockwise with respect to the Galactic plane (Figure~\ref{fig:polar_vector_scanam}). There are no obvious structures in the {\PILOT} total intensity maps that correspond to this characteristic orientation.}
    \item{Within our observed field, the POS magnetic field structure tends to become more parallel to the Galactic Plane in regions with high $240\,\mic$ brightness, such as the well-known  Sgr B2, 20\,\kms\ and 50\,\kms clouds (Figure~\ref{fig:l0_meanpsi_s240}).}
    \item {The polarization angles measured by {\PILOT} are in good agreement with the polarization angles measured by {\Planck} at 217 and 353\,GHz across the same region. The {\PILOT} data shows poorer agreement with the distribution of polarization angles measured by {\Planck} at lower frequencies, which is likely due to the impact of contamination by the CO line and synchrotron emission in in the Planck 70, 100 and 143\,GHz channels.}
    \item{The regularity of the POS magnetic field orientation and the low polarization fraction in the Galactic Center region suggest that the magnetic field is strong with a 3D geometry that is mostly oriented 15$\,^\circ$ with respect to the line-of-sight towards the \GC. Assuming equipartition between the magnetic pressure and ram pressure, we obtain magnetic field strengths estimates as high as a few mG for several CMZ molecular clouds (see Table~\ref{tbl:bfield_estimates}). }
   \item {We find no evidence for a connection between the orientation of the POS magnetic field structure inferred from the {\PILOT} observations and the twisted ring structure identified in the CMZ by \citet{Molinari_etal2011}.} 
   \item {
   We find broadly good agreement between the POS magnetic field orientations observed by {\PILOT} at $240\,\mic$ and by SCUBAPOL on the JCMT at $450\,\mic$ for a subfield centred on the the 50\kms molecular cloud. This suggests that the homogeneity of the POS magnetic field persists to smaller scales within the 50\kms molecular cloud region.}
\end{itemize}

\begin{appendix}

\section{Simplified Galactic Model}
\label{sec:gal_model_annex}
\label{sec:jbdmodel}

We model the Galactic plane as follows, inspired from models such as that of \cite{2013MNRAS.431..683J}. The dust density distribution is composed of an axisymmetric exponential profile, plus four spiral arms with Gaussian profiles in their cross sections. The magnetic field is the sum of an ordered component and a turbulent part, with amplitude proportional to the square root of the dust density (\cite{2010ApJ...725..466C}, \cite{2012ARA&A..50...29C}). The ordered part is spiral and the turbulent one is modeled as a Gaussian random field, isotropic and homogeneous, with its three components not correlated among them. Thus, all components have the same power spectrum, taken to be a power law with spectral index $\alpha$ in the range $[k_\mathrm{min},k_\mathrm{max}]$ and zero otherwise. A typical coherence length for this turbulent part is obtained by taking $k_\mathrm{min} = 10\,\mathrm{kpc}^{-1}$ (cf \cite{haverkorn&s_13}  and references therein). The precise value of the dissipation scale does not matter for our purpose as long as it is small enough, and we choose $k_\mathrm{max} = 1000\,\mathrm{kpc}^{-1}$. As for the index, we consider a Kolmogorov-like turbulence for which $\alpha = -11/3$.
We generate the turbulent magnetic field along a given line-of-sight as three independent Gaussian realizations (one for each component), while the values of that field along the adjacent lines-of-sight are constrained by the fact that the total three dimensional field has to be divertgentless. In practice, we do not need to explicit the precise values along these adjacent lines-of-sight since the following calculation is performed one line-of-sight at a time, but we note here that we indeed have enough freedom left in the construction of our field for it to be physical (i.e. divertgentless).

We compute the polarization fraction $p$ as follows. In principle it may be obtained using the expressions (5) to (7) of \cite{2015A&A...576A.105P}. However, for the purpose of this simple estimate, we note that to a first approximation we may neglect the optical depth effects and take $e^{\tau_\nu} = 1$ in the $I$, $Q$ and $U$ integrals for the frequency of interest (\cite{1996A&A...312..256B}), and that the source function $S_\nu$ and $p_0$ are constant along the line of sight. Therefore we have
\begin{equation}
{\footnotesize
p(s) = p_0 \frac{\sqrt{\left(\int_0^s n_d \cos (2 \phi) \cos^2 \gamma ds'\right)^2 + \left(\int_0^s n_d \sin (2 \phi) \cos^2 \gamma ds'\right)^2}}{\int_0^s n_d \left[1 - p_0 \left(\cos^2 \gamma - \frac{2}{3}\right)\right] ds'}
}
\end{equation}
where we use the same notations as in the aforementioned {\Planck} paper, $s$ is the distance along the line of sight, with origin at the Sun, and we take $p_0=0.2$. Note that with this definition $p$ can be greater than $p_0$.

\begin{figure}[!hb]
\centering
\includegraphics[width=0.4\textwidth]{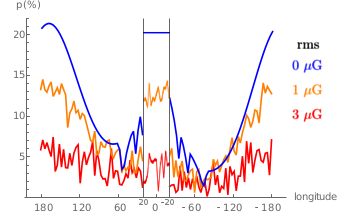}
\includegraphics[width=0.45\textwidth]{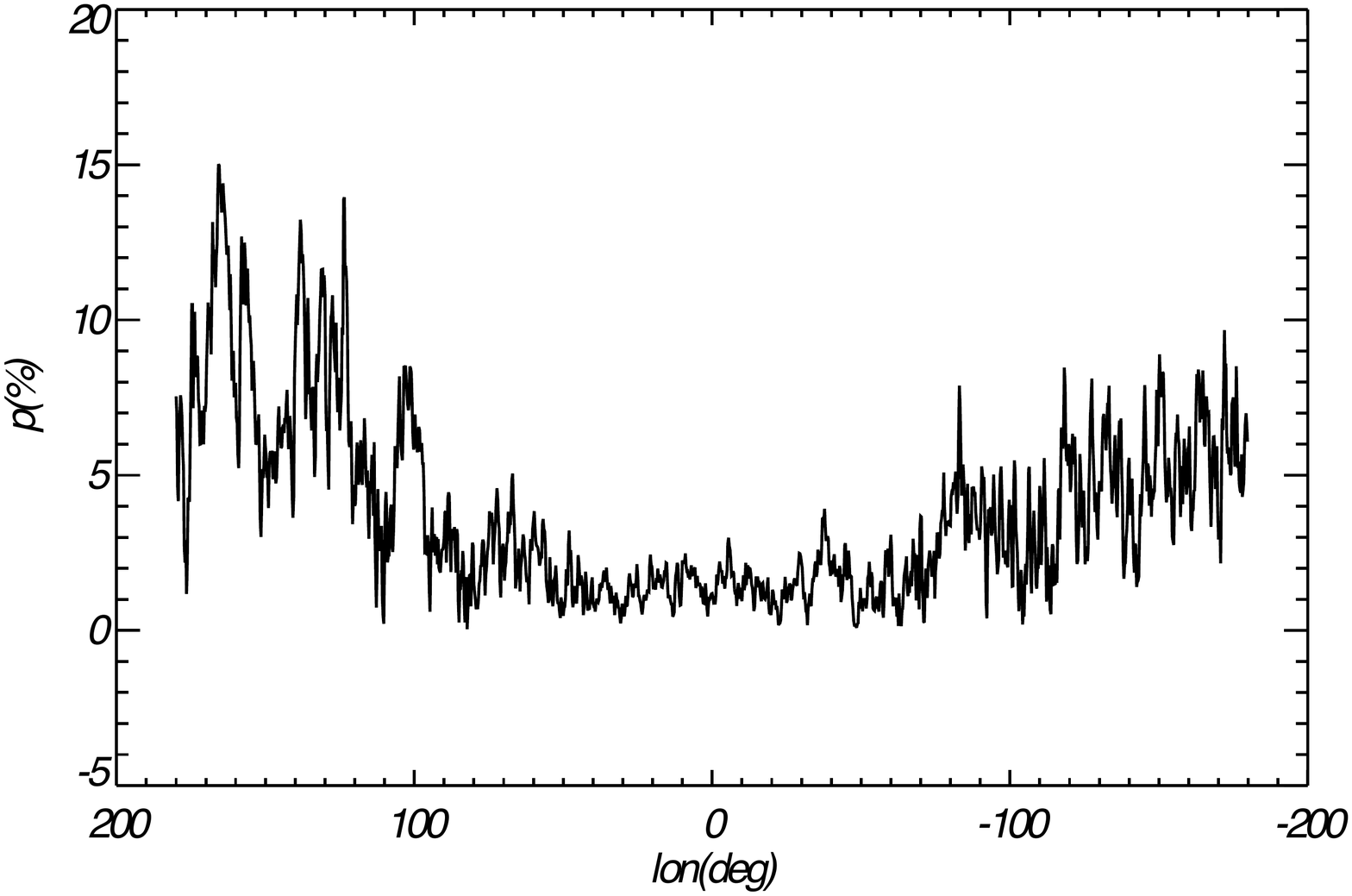}
\caption{Top: Polarization degree as a function of longitude in our model. The curves correspond to single realizations of the turbulent field, but are representative of the typical behaviour. Bottom: Measured values with Planck, using the modified asymptotic estimator \citep{Plaszczynski+2014} at a resolution of 15\,arcmin.}
\label{fig:pModelVsObs}
\end{figure}

We proceed in two steps. First, we consider the longitudes beyond the central region ($\ell \in [-180\,^\circ,-20\,^\circ] \cup [20\,^\circ,180\,^\circ]$) in order to evaluate how turbulent the magnetic field is, inside and between the spiral arms. In the top of figure \ref{fig:pModelVsObs} we show how in our model the polarization fraction depends on the longitude for three limiting cases: the blue curve corresponds to a switched off turbulence, orange corresponds to equipartition between the ordered and turbulent parts, and in red the root mean square of the turbulent part is three times larger than the ordered part. Comparing these results to the observed signal (bottom of that same figure) indicates that the turbulent component is important, with a typical root mean square of 2 to 3\,$\mu$G. Secondly, we focus on the longitudes $\ell \in [-20\,^\circ,20\,^\circ]$, and estimate the contribution to $p$ of everything along these lines of sight, except the bulge and the CMZ. To do so, we take a vanishing dust density inside the molecular ring, and compute $p$ using the estimated properties of turbulence in the disk from the first step. The results are plotted in figure \ref{fig:pModelVsObs}. Naturally, the stronger the turbulence, the lower $p$ is. But $p$ does not reach values as low as the measured ones. Indeed, the observed $p$ is of typically $1\,\%$ throughout that central region, while even in the case of a strong turbulence (red curve) $p$ is around 4\,$\%$. Therefore the central part has a non negligible contribution to the observed signal.

\end{appendix}

\begin{acknowledgements}
This work was supported by the Programme National ``Physique et
Chimie du Milieu Interstellaire'' (PCMI) of CNRS/INSU with INC/INP co-funded by CEA and CNES.

\end{acknowledgements}

\bibliographystyle{aa}

\bibliography{pilot_L0}

\end{document}

%% file: journals_defs.tex
\def\apjl{ApJ}%
\def\apjs{ApJS}%
\def\ao{Appl.~Opt.}%
\def\aap{A\&A}%

%% file: polar_definitions.tex
\newcommand{\StokesI}{I}                    
\newcommand{\StokesQ}{Q}                    
\newcommand{\StokesU}{U}                    
\newcommand{\polfrac}{p}                    
\newcommand{\polang}{\psi}                  
\newcommand{\DeltaAng}{\mathcal{S}} 

